\documentclass[
 reprint,
 onecolumn,
 superscriptaddress,
 preprintnumbers,
 nofootinbib,
 amsmath,amssymb,
 aps,
 floatfix,
]{revtex4-2}
\usepackage{graphicx, amsmath,amssymb,subcaption}
\usepackage{color,bm}
\usepackage{makecell}

\usepackage[colorlinks,linkcolor=blue]{hyperref}
\usepackage[titletoc]{appendix}
\usepackage{float}
\begin{document}

%\preprint{}

\title{On the energy flow  of  $\lambda$ in Ho\v{r}ava-Lifshitz cosmology}

\author{Ewa Czuchry}
\email[]{ewa.czuchry@ncbj.gov.pl}
\affiliation{National Centre for Nuclear Research, Pasteura 7, Warsaw, Poland}

\author{Nils A. Nilsson}
\email{nils.nilsson@obspm.fr}

\affiliation{SYRTE, Observatoire de Paris-PSL, Sorbonne Universit\'e, CNRS UMR8630, LNE, 61 avenue del’Observatoire, 75014 Paris, France}

\date{\today}

\begin{abstract}
 Ho\v{r}ava-Lifshitz gravity has been proposed as a ghost-free quantum gravity model candidate with an anisotropic UV-scaling between space and time. We present here a cosmological background analysis of two different formulations of the theory, with particular focus on the running of the parameter $\lambda$. Using a large dataset consisting of Cosmic Microwave Background data from {\it Planck}, Pantheon+ supernovae catalogue, SH0ES Cepheid variable stars, Baryon acoustic oscillations (BAO), Cosmic Chronometers, and gamma-ray bursts (GRB), we arrive at new bounds on the cosmological parameters, in particular $\lambda$, which describes deviation from classical general relativity. 
    For the detailed balance scenario we arrive at the bound $\lambda=1.02726\pm0.00012$, and for beyond detailed balance the limit reads $\lambda=0.9949^{+0.0045}_{-0.0046}$. We also study the influence of different data sets and priors, and we find that removing low-redshift data generally moves $\lambda$ closer towards UV values, whilst simultaneously widening the error bars. In the detailed balance scenario, this effect is more noticeable, and $\lambda$ takes on values that are significantly below unity, which corresponds to the infrared limit of the theory. 
    
\end{abstract}

%\keywords{}

\maketitle

\section{Introduction}

General Relativity (GR), one of the most successful physical theories, is widely known to be non-renormalizable. Therefore, its application to very small distances and high energies, such as at the very early Universe is expected to be inconsistent. In light of this, a different theory of gravity was recently proposed by Ho\v{r}ava \cite{Horava:2009if} in order  to capture the quantum effects of the gravitational field in the early stages of the Universe. This theory is a fascinating proposal of modified gravity equipped with an anisotropic scaling
 at the Planck scale given in term of a critical Lifshitz exponent which contributed to the name  Ho\v{r}ava-Lifshitz (HL) gravity. This Lifshitz scaling in the ultraviolet (UV) regime inevitably breaks Lorentz invariance explicitly\footnote{See for example \cite{Bluhm:2014oua} for a discussion of explicit and spontaneous symmetry-breaking in gravity.}.

By giving up on the idea of space-time invariance under four-dimensional diffeomorphisms it becomes possible to add higher spatial-derivative terms to the Lagrangian in a way that makes the theory renormalizable at high energies. The classical formulation by Ho\v{r}ava \cite{Horava:2009if}  included in the potential part of the action only terms  which may  be derived from a superpotential (the so called "detailed balance condition"), thus limiting the number of its
terms to those quadratic in curvature and with independent couplings. Subsequent formulations, the so called Sotiriou-Visser-Weinfurtner (SVW) generalisation  \cite{Sotiriou:2009bx} included, relaxed this condition and included cubic terms in curvature in the gravitational action.

The main feature distinguishing HL from GR is the explicit lack of invariance under four-dimensional diffeomorphisms; in HL, there exists  an ultraviolet fixed point, where time and space scale differently, and so HL is naturally expressed using the Arnowitt-Deser-Misner (ADM) 3+1 formulation \cite{Arnowitt:1962hi}.
The most straightforward option for a group which can accommodate this property is the set of foliation-preserving diffeomorphisms, which includes time reparametrizations and three-dimensional spatial diffeomorphisms. Under this symmetry group, the kinetic term of the action acquires an additional coupling denoted by $\lambda$. In GR there is no need for that coupling as the very specific linear combination of curvature terms contained within the classical Hilbert-Einstein action remains unchanged under general four-dimensional diffeomorphisms. In HL theory the limit $\lambda\rightarrow1$ is supposed to recover classical general relativity.
This should in theory provide its low-energy limit in a straightforward manner, but achieving this limit presents some issues: on one hand, the critical Lifshitz exponent $z$ and the $(3+1)$-foliation parameter $\lambda$ are the parameters of HL theory, where $\lambda$ is also  associated with a restricted foliation corresponding to $z=1$ in the Lifshitz scaling. As $z$ approaches unity in the low-energy limit, it is also necessary that $\lambda\rightarrow1$ and consequently that Lorentz invariance is restored. Additionally, the familiar  ADM foliation of GR is also recovered. Because the coupling constant $\lambda$ disrupts the general covariance in the presence of space-time diffeomorphisms, it is thought that values of $\lambda$ different from $1$ will cause the model to deviate from GR.
On the other hand, due to the reduced symmetry, the action of HL gravity \cite{Horava:2009if} differs from the classical Einstein-Hilbert action even in the infrared limit, where  only terms up to second order in spatial derivatives are considered. 
Even though the IR limit differs from GR, it is supposed that one can still obtain it as an approximate solution at long distances; however, it seems that this connection with GR is problematic, since in order to obtain it, it is necessary to disregard the higher order derivative terms in the action. Additionally, the GR limit presents issues because HL contains an extra degree of freedom 
present as a scalar mode, which causes a significant challenge in terms of the credibility of the theory \cite{Sotiriou:2009bx,Bogdanos:2009uj,Koyama:2009hc}. 

In recent years, significant work has been done on HL gravity and its cosmological aspects (see \cite{Wang:2017brl} for an extensive review), including several early extensions attempting to cure some of its initial problems~\cite{Audren:2014hza,Blas:2009qj,Blas:2011zd,Calcagni:2009ar,Colombo:2015yha,Czuchry:2009hz,Czuchry:2010vx,Dutta:2009jn,Dutta:2010jh,Frusciante:2015maa,Kiritsis:2009sh,Lu:2009em,Saridakis:2009bv,Sotiriou:2009bx},   {as well as studies of black holes \cite{Park:2009zr}, dark energy \cite{Park:2009zra}, the black-hole shadow \cite{Li:2021riw,Jusufi:2022ava}, gravitational waves \cite{Gong:2021jgg,EmirGumrukcuoglu:2017cfa}, and many more (see \cite{Wang:2017brl})}. There has been also a lot of discussion on its problems and inconsistencies (see e.g.  \cite{Sotiriou:2009bx,Blas:2009qj, Sotiriou:2010wn}) such as the existence of a parity violating term \cite{Sotiriou:2009bx},
ghost instabilities and problems with strong coupling at very low energies \cite{Blas:2009yd, Charmousis:2009tc}, wrong sign and very large value of the cosmological constant \cite{Vernieri:2011aa, Appignani:2009dy}, problems with power counting renormalization of the scalar mode propagation \cite{Vernieri:2015uma,Colombo:2015yha}. Some authors state that most of the  shortcomings can
be cured by adding additional terms to the superpotential, or relaxing the projectability condition which assumes that the lapse function $ N $ depends only
on time ${ N }={ N }(t)$, resulting in its {\it non-projectable} version.  Additionally, given that the HL theory is an extension of GR in the high energy regime, it is unsurprising that also its quantum gravity aspects have sparked considerable interest \cite{Bertolami:2011ka,Christodoulakis:2011np,Pitelli:2012sj,Vakili:2013wc,Obregon:2012bt,Benedetti:2014dra,Garcia-Compean:2021vcy}.

In the first "projectable" versions (${ N }={ N }(t)$), including the detailed balance formulation (and further the Sotiriou-Visser-Weinfurtner extension \cite{Sotiriou:2009bx} with the relaxing of the detailed balance condition), application of the action to the maximally symmetric Friedmann-Lemaitre-Robertson-Walker (FLRW) metrics results in the instability of the extra scalar mode. In general, $\lambda$ runs with energy  and has  the three IR fixed
points:  $\lambda=1/3$, $\lambda=1$  and $\lambda=\infty$, and the unstable quantum behaviour of the scalar mode results in a ghost, appearing in the range $1/3< \lambda <1$, whereas it is only classically unstable in other ranges: $\lambda <1 /3$ or $\lambda >1$ \cite{Sotiriou:2009bx,Bogdanos:2009uj,Koyama:2009hc}. The classical instability in the UV range, when we expect quantum effects, is rather of lesser importance.

The "non-projectable" version of HL, the so called called "healthy extension" \cite{Blas:2009qj} was supposed to  eliminate this unphysical scalar degree of freedom; nevertheless, there has been a controversy regarding the actual health benefits of these extensions, as was demonstrated in \cite{Papazoglou:2009fj}. There, it was shown that difficulties with strong coupling appear in the low-energy limit, and that the limit itself is problematic, since cubic terms in the action diverge when $\lambda\rightarrow 1$. 
As a rebuttal, the authors of the original paper have put forth the argument that the strong coupling scale could potentially surpass the cutoff point for derivative expansions \cite{Blas:2009ck}, which would introduce a substantially lower scale of Lorentz violation.

 Besides theoretical considerations there have been several attempts to put observational bounds on $\lambda$ in addition to other HL gravity parameters.
The first works on observational bounds on HL gravity parameters such as~\cite{Dutta:2009jn}   {and our previous paper \cite{Nilsson:2018knn}}  has {manually} set $\lambda$ to unity in the analysis   {by considering the data to be too far in the IR compared to the scale where this parameter can be expected to run.  In \cite{DiValentino:2022eot} and the work by one us \cite{Nilsson:2021ute}, the authors considered a constant $\lambda$ and reabsorbed it into other parameters. It is however worth mentioning that \cite{Dutta:2010jh} obtained the first constraints on $\lambda$ from cosmological data, reporting $|\lambda-1|\lesssim 0.02$,   {while one of us }\cite{Nilsson:2019bxv} obtained $0.95\leq\lambda\leq1.16$ by exploiting bounds on the Hubble parameter tension, and \cite{Xu:2018fag} investigated the effects of a $\lambda$ phase transition in AdS black holes.}

There are also other studies \cite{Frusciante:2015maa} providing bounds on the different parameters of HL cosmology  based on the so-called healthy extension of HL \cite{Blas:2009qj}, where the authors employ an effective-field theory and also consider linear perturbations around the background. This study is however based on a flat model, whereas there is ongoing discussion on the  possible non-zero value of  the curvature parameter both in $\Lambda$CDM model \cite{Handley:2019tkm} as its alternatives including HL cosmology.

 {Therefore, a natural question arises: how does the cosmological behaviour of Ho\v{r}ava-Lifshitz gravity change when $\lambda$ is allowed to vary, since it is supposed to control the breaking of Lorentz invariance, and it might flow to unity deep in the IR. When we allow $\lambda$ to vary we can set bounds on the ``level`` of Lorentz violation at the energy scales available to us through data.  {This was  done in \cite{Dutta:2010jh} 
but this analysis simply took into account all the available then data and did not investigate the influence of different energy scales on the values of $\lambda$, which is what we do in the present paper. Moreover, in this manuscript we use 
 a larger and more updated dataset with  an additional number of high-energy sources such as gamma-ray bursts, which may have big influence on numerical results}.

 {In this paper we analyze the influence of including/excluding high/low energy observational data on the values of $\lambda$ and perform a detailed examination regarding how removing low-redshift data impact the obtained bounds on this parameter. An intriguing matter comes to light, whether higher redshift data could push $\lambda$ further from the classical limit at the value of unity, as would be expected of HL or similar extensions of Einstein gravity (with additional terms in the action). In this paper we perform the analysis of including different data sets and priors on the value of $\lambda$ in two {\it projectable} versions of HL, one with imposed detailed balance condition and one with this condition relaxed.}

This paper is organized as follows: We first give a brief overview of the theory of HL cosmology in both scenarios under consideration. We then go on to describe how we rewrite the equations to allow for numerical analysis. Finally, we present our results and discuss them.

\section{Ho\v{r}ava-Lifshitz cosmology}
Here we briefly outline the equations governing cosmological evolution in Ho\v{r}ava-Lifshitz gravity~\cite{Calcagni:2009ar,Kiritsis:2009sh}. For a more in-depth discussion we refer to our recent paper \cite{Nilsson:2018knn}, and also to \cite{Wang:2017brl}.

Ho\v{r}ava-Lifshitz gravity offers the possibility of creating a theory of gravity which remains finite and well defined at high energies, while also reproducing GR in the classical regime. This is achieved by introducing a fixed point in the UV region of the renormalization group, where time and space behave differently under scale transformations. Specifically, the scaling relations at the Planck scale should satisfy $t\,\rightarrow\,b^{-z}t$, $x^i\,\rightarrow\,b^{-1}x^i$, $i=1,2,3$, where  $b$ is a scaling parameter and $z$ is the critical Lifshitz exponent which characterizes the fixed point. As usual, $t$ stands for time and $x^i$ for spatial coordinates. Different models can be identified by specific choices of $z$; for a pure gravity theory in $d$ spatial dimensions which is invariant under foliation-preserving diffeomorphisms and is power-counting renormalizable, $z$ must be greater than or equal to $d$,  equivalently in 4-dimensional physical space-time to  $z\geq4$~\cite{Wang:2017brl}. In order to restore Lorentz invariance, $z$ needs to be set to
unity.

Due to the anisotropic scaling present in the theory, it is useful to write down the metric using the ADM decomposition:
\begin{equation}\label{eq:metric}
    ds^2=-N^2dt^2+g_{ij}(dx^i+N^idt)(dx^j+N^jdt),
\end{equation}
where $N$ and $N^j$ are the lapse function and shift vector, respectively, and $g_{ij}$ is the spatial metric ($i,j=1,2,3$). Given this, we can express the most general form of the theory as:
\begin{equation}\label{eq:action1}
    S=\int d^3xdtN\sqrt{g}\left[K^{ij}K_{ij}-\lambda K^2-\mathcal{V}(g_{ij})\right],
\end{equation}
where $g$ is the determinant of the spatial metric, $\lambda$ is the running coupling and $\mathcal{V}$ is a potential. $K_{ij}$ represents the extrinsic curvature. The potential term contains only dimension 4 and 6 operators which can be constructed from the spatial metric $g_{ij}$.
The square $K^{ij}K_{ij} $  and its trace-squared $K^2$ are individually invariant under the reduced symmetry group, however for $\lambda=1$ the full kinetic term 
$K^{ij}K_{ij}- K^2$ acquires invariance under four-diffeomorphisms.

\subsection{Detailed balance (DB)}
The detailed-balance condition is a way to reduce the number of terms in the action~(\ref{eq:action1}) by assuming that it should be possible to derive $\mathcal{V}$ from a superpotential W~\cite{Horava:2009uw,Vernieri:2011aa}:
\begin{equation}\label{eq:superpotential}
    \mathcal{V}=E^{ij}\mathcal{G}_{ijkl}E^{kl}, \quad E^{ij}=\frac{1}{\sqrt{g}}\frac{\delta W}{\delta g_{ij}}, \quad \mathcal{G}^{ijkl}=\frac{1}{2}\left(g^{ik}g^{jl}+g^{il}g^{jk}\right)- \lambda g^{ij}g^{kl}.
\end{equation}
which for $\lambda = 1$ reduces to the standard Wheeler-DeWitt metric. From this the most general action can be written as:
\begin{align}\label{eq:action_det}
        S_{db} = \int \text{d}^3x\,\text{d}t \sqrt{g}N&\Bigg[\frac{2}{\kappa^2}\left(K_{ij}K^{ij}-\lambda K^2\right)+\frac{\kappa^2}{2\omega^4}C_{ij}C^{ij}-\frac{\kappa^2\mu}{2\omega^2}\frac{\epsilon^{ijk}}{\sqrt{g}}R_{il}\nabla_jR^l_k \\&+\frac{\kappa^2\mu^2}{8}R_{ij}R^{ij}+\frac{\kappa^2\mu^2}{8(1-3\lambda)}\left(\frac{1-4\lambda}{4}R^2+\Lambda R -3\Lambda^2\right)\Bigg],
    \end{align}
where $C^{ij}$ is the Cotton tensor, $\epsilon^{ijk}$ is the totally antisymmetric tensor, and the parameters $\kappa$, $\omega$, and $\mu$ have mass dimension $-1$,$0$, and $1$, respectively. This action has been obtained from (\ref{eq:action1}) by analytic continuation of the parameters $\mu \mapsto i\mu$ and $\omega^2 \mapsto -i\omega^2$, which enables positive values of the bare cosmological constant $\Lambda$, which does not happen in the original formulation.

As mentioned in the Introduction, the coupling constant $\lambda$ runs with energy  {in a complicated manner} and may reach one of the IR fixed points $\{1/3; 1; \infty\}$~\cite{Horava:2009uw}.  From the point of view of scalar perturbations it was pointed out in \cite{Sotiriou:2009bx} that the range $1/3 <\lambda <1$
 leads to instabilities. Unfortunately, also the flow interval of $\lambda$ between the UV and IR regimes exhibits an unstable quantum behaviour of the scalar mode \cite{Sotiriou:2009bx,Bogdanos:2009uj,Koyama:2009hc}.
 An effort was made to address this problem by proposing, as suggested in \cite{Bogdanos:2009uj}, the enforcement of an analytic continuation of a particular form; however, upon performing the analytic continuation as described in \cite{Bogdanos:2009uj}, it becomes evident that the desirable UV behaviour is compromised and consequently, instabilities resurface at high energies.
 
 If we exclude the use of analytic continuation, the only relevant scenario which permits a potential flow towards General Relativity (at $\lambda=1$) is when $\lambda\geq1$, as the region for  values of $\lambda$ less than or equal to 1/3 is separated by a problematic interval. 
However, in this paper we \emph{have not imposed this condition on $\lambda$}. Rather, we
left it as free as possible, the idea being that certain formulations of HL gravity may be ruled out by the model preferring unphysical values of $\lambda$ when confronted with data.\footnote{In the initial numerical results, we noted that whilst $\lambda$ does indeed take on values smaller than unity, it never gets close to $1/3$ (which is parameter singularity), and we therefore reintroduced the $\lambda>1/3$ prior to save computation time, without loss of generality.} Also the idea was to fit the observational data without {\it a priori} assumption, allowing further theoretical considerations based on obtained values. For completeness, we also investigate the case when $\lambda\geq1$.

  {Other formulations other than the one presented above exist; for example, one can improve the low-energy behaviour of the theory by means of an IR-modification parameter which for example allows for a consistent Minkowski vacuum, as was done in \cite{Kehagias:2009is,Park:2009zra}. Such an IR-modification constitutes a so-called soft breaking of the detailed-balance condition, and we do indeed see these features when breaking this condition in this paper. Modifications can also be introduced in the UV limit in the form of fourth, fifth, and sixth order derivative terms; a UV detailed-balance condition is normally in place which cancels these terms, but one may break this condition softly, and this is in fact necessary in order to obtain scale-invariant scalar perturbations, as was shown in \cite{Shin:2017ott,Kobayashi:2010eh}. In this paper, we use the formulation from \cite{Horava:2009if} and \cite{Charmousis:2009tc} which possesses desirable features when relaxing the detailed-balance condition; for completeness, we also study the case when this condition is in place.}

We populate our model with the canonical matter and radiation fields represented by the energy densities (and pressures) $\rho_m$ ($p_m$) and $\rho_r$ ($p_r$), both of which are subject to the continuity equation $\dot{\rho} + 3H(\rho+p)=0$. Here, an overdot represents a time derivative. Moreover, we use the projectability condition \cite{Horava:2009uw} $N=N(t)$, and we use the standard FLRW line element $g_{ij} = a(t)^2\gamma_{ij}$, $N^i=0$, where $\gamma_{ij}$ is the maximally symmetric constant
curvature metric.
It is important to remember that in theories which violate Lorentz symmetry the gravitational constant appearing in the gravitational action $G_{\text{grav}}$ generally does \emph{not} coincide with the one which appears in the Friedmann equations, $G_{\text{cosmo}}$~\cite{Carroll:2004ai,Dutta:2010jh}. This could in principle be used to set bounds on $\lambda$ (as was done in \cite{Nilsson:2019bxv}), but we will not adopt this approach in this paper.

Varying the action (\ref{eq:action_det}) w.r.t $N$ and $a$ we arrive at the Friedmann equations for the detailed balance scenario:
\begin{equation}\label{eq:friedmann1}
    \left(\frac{\dot{a}}{a}\right)^2 = \frac{\kappa^2}{6(3\lambda-1)}\left[\rho_m+\rho_r\right]+\frac{\kappa^2}{6(3\lambda-1)}\left[\frac{3\kappa^2\mu^2 K^2}{8(3\lambda-1)a^4}+\frac{3\kappa^2\mu^2\Lambda^2}{8(3\lambda-1)}\right]-\frac{\kappa^4\mu^2\Lambda K}{8(3\lambda-1)^2a^2},
\end{equation}
\begin{equation}\label{eq:friedmann2}
    \frac{\text{d}}{\text{d}t}\frac{\dot{a}}{a} + \frac{3}{2}\left(\frac{\dot{a}}{a}\right)^2 = -\frac{\kappa^2}{4(3\lambda-1)}[p_m+p_r]-
     \frac{\kappa^2}{4(3\lambda-1)}\left[\frac{\kappa^2\mu^2 K^2}{8(3\lambda-1)a^4}-\frac{3\kappa^2\mu^2\Lambda^2}{8(3\lambda-1)} \right]-\frac{\kappa^4\mu^2\Lambda K}{16(3\lambda-1)^2a^2}.
\end{equation}
We can therefore define $G_{\text{cosmo}}=\kappa^2/(3\lambda-1)$ and $\kappa^4\mu^2\Lambda= 8(3\lambda-1)^2$ by requiring that (\ref{eq:friedmann1},\ref{eq:friedmann2}) coincide with the standard Friedmann equations. Clearly, when Lorentz invariance is restored, $\lambda$ is set to unity and $G_{\text{grav}}=G_{\text{cosmo}}$.
Under detailed balance, and using the units $8\pi G_{\text{grav}} = 1$, we are lead to $\kappa^2 = 4$, $\mu^2\Lambda = 2$, and by introducing the standard density parameters we arrive at the Friedmann equation suitable for our analysis:
\begin{equation}\label{eq:friedmann_det2}
    H^2 = H_0^2\left[\frac{2}{3\lambda-1}\Large(\Omega_{m0}(1+z)^3+\Omega_{r0}(1+z^4)\Large)+\Omega_{K0}(1+z)^2+\omega+\frac{\Omega_{K0}^2}{4\omega}(1+z)^4\right],
\end{equation}
where $H$ denotes the Hubble parameter and the subscript $0$ indicates the value as measured today. We have also introduced a parameter $\omega = \Lambda/(2H_0^2)$\footnote{Which is not to be confused with the IR-modification parameter, which is usually also denoted by $\omega$.}. A characteristic feature of HL theory is the appearance of a dark radiation term in (\ref{eq:friedmann_det2}): $\Omega_{k0}^2/4\omega$, and we can express this in terms of the effective number of neutrino species $\Delta N_{\rm eff}$ present during the BBN epoque (See~\cite{Dutta:2009jn,Dutta:2010jh,Nilsson:2018knn}) as
$\Omega_{k0}^2/4\omega=0.13424\Delta N_{\rm eff} \Omega_{r0}$. We can also obtain a constraint from the $z=0$ limit, where $H|_{z=0} = H_0$, which reads: $(1-\Omega_{k0}-\omega-\Omega_{k0}^2/(4\omega))(3\lambda-1)/2=\Omega_{m0}+\Omega_{r0}$. We abbreviate the Detailed Balance case as DB.

\subsection{Beyond detailed balance (BDB)}
There has been an ongoing discussion in the literature whether the detailed balance condition is too restrictive~\cite{Calcagni:2009ar,Kiritsis:2009sh,Wang:2017brl}, or if relaxing it can cure the theory from the quantum instabilities of the scalar mode \cite{Blas:2009qj,Koyama:2009hc,Kiritsis:2009vz,Henneaux:2009zb,Chen:2009vu}. On one hand the detailed-balance condition introduces a superpotential, which might simplify the quantization process significantly; on the other hand, this condition is not fundamental, but helps in {\it e.g.,} reducing the number of independent couplings in the theory.
 
Therefore, we may safely consider the Sotiriou-Visser-Weinfurtner (SVW) \cite{Sotiriou:2009bx}   generalization with this condition relaxed, after which the potential $\mathcal{V}$ contains additional terms. In this scenario the action includes quantities not only up to
quadratic in curvature, as in the original HL formulation, but also cubic ones, while suppressing parity violating terms. Applying the generalized action to the maximally symmetric constant curvature metric results in the following analogue of the Friedmann equations~\cite{Charmousis:2009tc,Sotiriou:2010wn}:
\begin{equation}\label{eq:friedmann3}
    \left(\frac{\dot{a}}{a}\right)^2 = \frac{2\sigma_0}{3\lambda-1}(\rho_m+\rho_r)+\frac{2}{3\lambda-1}\left[\frac{\Lambda}{2}+\frac{\sigma_3K^2}{6a^4}+\frac{\sigma_4K}{6a^6}\right]+\frac{\sigma_2}{3(3\lambda-1)}\frac{K}{a^2},
\end{equation}
\begin{equation}
    \frac{\text{d}}{\text{d}t}\frac{\dot{a}}{a}+\frac{3}{2}\left(\frac{\dot{a}}{a}\right)^2 = -\frac{3\sigma_0}{3\lambda-1}\frac{\rho_r}{3}-\frac{3}{3\lambda-1}\left[-\frac{\Lambda}{2}+\frac{\sigma_3K^2}{18a^4}+\frac{\sigma_4K}{6a^6}\right]+\frac{\sigma_2}{6(3\lambda-1)}\frac{K}{a^2},
\end{equation}
where $\sigma_i$ are arbitrary constants. As in the detailed balance scenario we find $G_\text{cosmo}=6\sigma_0/(8\pi(3\lambda-1))$, where $\sigma_0=\kappa^2/12$. Using the same procedure as for detailed balance in units where $8\pi G_\text{grav}=1$  we rewrite (\ref{eq:friedmann3}) to read:
\begin{equation}\label{eq:friedmann4}
    H^2=H_0^2=\left[\frac{2}{3\lambda-1}\Big(\Omega_{m0}(1+z)^3+\Omega_{r0}(1+z)^4+\omega_1+\omega_3(1+z)^4+\omega_4(1+z)^6\Big]+\Omega_{k0}(1+z)^2\right],
\end{equation}
where we have, for convenience, introduced the following dimensionless parameters, $\omega_1=\sigma_1/(6H_0^2)$, $\omega_3=\sigma_3H_0^2\Omega_{k0}^2/6$, $\omega_4=-\sigma_4\Omega_{k0}/6$. Additionally, we impose $\omega_4>0$ in order for the Hubble parameter to be real for all $z$. Moreover, we can extract a constraint from the $z=0$ limit of (\ref{eq:friedmann4}), which then reads
$(1-\Omega_{k0})(3\lambda-1/2=\Omega_{m0}+\Omega_{r0}+\omega_1+\omega_3+\omega_4$. Much like for detailed balance, we can eliminate another parameter through constraints by considering that the $\omega_4$ term corresponds to a quintessence-like kinetic field. Therefore it is shown in \cite{Dutta:2009jn} and references therein that we get the following constraint at the time of BBN: $\omega_3 = 0.13424\Delta N_{\rm eff} \Omega_{r0}-\omega_4(1+z_{BBN})^2$. Here $z_{BBN} \approx 4\cdot 10^4$
is the redshift at BBN. We abbreviate the Beyond Detailed Balance case as BDB.

\section{Constraints from cosmological data}
\subsection{The data}
In order to carry out the parameter estimation we used a Markov-Chain Monte Carlo (MCMC) method with a large cosmological data set. Since we are interested in the running of $\lambda$ with energy, we include here a short discussion of the different energy levels of the cosmological probes we use. It is not a priori clear whether (or how) the running of $\lambda$ depends on this energy.

Starting at high redshift (early times), our probes are the Cosmic Microwave Background (CMB) and Baryon Acoustic Oscillations (BAO). The CMB originates from the
surface of last scattering at redshift $z_\star \sim 1100$, when the temperature of the Universe was around $T \geq 3000$K, corresponding to an energy of around $0.26$eV. This is complemented by BAO observations, which also originate from recombination and manifest as acoustic peaks in the CMB power spectrum; however, they also affect the distribution of local galaxies and are therefore sensitive to different parameters compared to the CMB whilst still being an early Universe probe. 

The astrophysical sources we used have different properties, in that they originate in the \emph{late} Universe, and whilst the emission energies are in general higher than CMB and BAO, they do not offer any information about the state of the Universe at early times. Supernovae Type 1a are used as standard rulers and outputs more energy than the rest of its host galaxy in a very short time. Their energy is less important in this context as they are used as distance rulers by fitting their spectra; however, they are
certainly high-energy events, emitting neutrinos with energies up to $\sim 40$ MeV having been detected~\cite{Janka:2017vlw}. The Cosmic Chronometer (CC) data set is based on passively evolving galaxies, and we only sample the expansion history with these objects. Last and most energetic, we also have gamma-ray bursts, which are the most
violent explosions in the known Universe; photons from gamma-ray bursts have been found to reach energies up to $96$ GeV~\cite{PIRON2016617}. For all details about the data sets and their implementation, see Appendix~\ref{app:data} and references therein.

\subsection{Results}
  {In order to derive constraints on the parameters we used (\ref{eq:friedmann_det2}) and its associated constraint equations in our MCMC analysis. As we are mainly interested in the running of $\lambda$ we present here only the marginalized posteriors of this parameter. We include all parameter fits in Table~\ref{tab:DB} and \ref{tab:BDB} which can be found in Appendix~\ref{app:tables}, where we are specifically interested in $\Omega_{k0}$ and $\Delta N_{\rm eff}$, since previous work (\cite{Dutta:2009jn,Dutta:2010jh,Nilsson:2018knn}) suggests that it is possible to differentiate between the two scenarios using these quantities.}

  {In our analysis we find that in the detailed-balance scenario using all available data, $\lambda$ takes the value $\lambda=1.02726\pm0.00012$ at $1\sigma$ confidence level, when not imposing the hard prior $\lambda\geq1$; its marginalized posterior is shown in Figure~\ref{fig:lambdaDBnoprior}\footnote{The MCMC chains which generated the posteriors passed the standard convergence criteria detailed in~\cite{Dunkley:2004sv}.}. After imposing the prior on $\lambda$, we instead obtain $\lambda=1.046\pm0.0023$, and note that this value is a few percent larger, but the $1\sigma$ error bars are one order of magnitude larger, which can be seen in Figure~\ref{fig:lambdaDBprior}.}
{In order to investigate the possible running of $\lambda$ with energy, we carry out the same analysis by systematically removing data in two steps: in the first step, we remove the Hubble parameter measurements from Cosmic Chronometers (CC), and in the second step we remove everything except the truly early-Universe probes, CMB and BAO. For $\lambda$, the results are displayed in Figure~\ref{fig:lambdas}, and we observe that the results are different depending on whether we impose the Detailed-Balance condition or not. First of all, $\lambda$ generally seems to take on higher values under Detailed Balance; for this case, removing CC data pushes $\lambda$ to take values close to unity, but further removing data moves it back up toward higher values, but with much larger error bars. For both DB and BDB, the CMB+BAO combination seems to give the largest errors by far; this is especially pronounced in DB (no prior) and BDB (prior). }

  {The case with the fewest assumptions and priors is the BDB (no prior) case shown in Figure~\ref{fig:lambdaBDBnoprior}, and the situation here is somewhat reversed compared to the other three; here, removing CC data produces a higher value of $\lambda$ than with the full dataset, which is in contrast to our other results. Also, the case with all data overlaps significantly with that of CMB+BAO, both of which have means firmly in the $\lambda<1$ region.}
{In almost all presented scenarios, in DB and BDB with the prior $\lambda\ge1$ imposed and in DB with no prior we observe that removing Hubble data from CC strongly pushes $\lambda$ close to its IR limit. Only in the BDB scenario with no imposed prior does removing that data result in $\lambda$ changing sign and increasing $|\lambda-1|$. Furthermore, removing Supernovae Type 1a  pushes $\lambda$ even further into UV, except in the BDB with no prior scenario.} 

{Another interesting quantity to analyse is the difference between $G_{\rm grav}$ and $G_{\rm cosmo}$ which can be inferred directly from $\lambda$ through the quantity $|G_{\rm cosmo}/G_{\rm grav}-1|$. Here, we find that for detailed balance, the difference between local $G$ and cosmological $G$ is in general smaller than $\sim5.8\%$, the largest discrepancy being in the case of all data when including a prior on $\lambda$, where we find $|G_{\rm cosmo}/G_{\rm grav}-1|=0.0574\pm0.0030$. For beyond detailed balance, this quantity takes on smaller values, generally $<2\%$, which is of course directly linked to the values obtained on $\lambda$.}

  {As in our previous work \cite{Nilsson:2018knn} we find here that $\Omega_{k0}$ is distinctly non-zero in the Detailed Balance formulation of HL cosmology, which is the same as was found in \cite{Nilsson:2021ute,DiValentino:2022eot}. We also briefly point out our results on the Hubble constant, which are somewhat different from those of previous background analysis \cite{Nilsson:2018knn,Nilsson:2021ute}, where values close to $h=0.71$ were found for the case of all data, which eased the Hubble parameter tension. In our present results, we find values closer to that of $\Lambda$CDM, with $h=0.6488\pm0.0012$ for DB and $h=0.6813\pm0.048$ for BDB, both with all data; the BDB value closely resembles that obtained by {\it Planck} ($h=0.6844\pm0.091$) using TE+lowE spectra whilst assuming a base $\Lambda$CDM model. On the other hand, the DB case takes values close to the {\it Planck} TT+TE+EE+lowE+lensing results when considering a $\Lambda$CDM model extended by allowing for a non-zero $\Delta N_{\rm eff}$ \cite{Planck:2018vyg}. A full analysis of the Hubble tension lies beyond the scope of this paper. }

\begin{figure}
        \centering
        \begin{subfigure}[b]{0.495\textwidth}
            \centering
            \includegraphics[width=\textwidth]{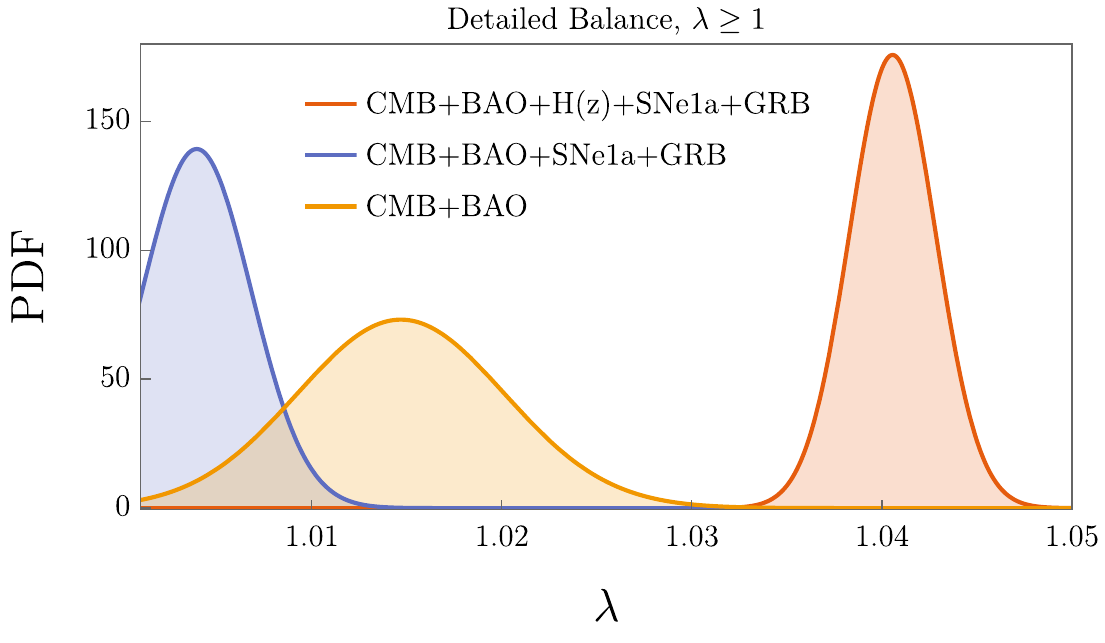}
            \caption[DB, prior]%
            {{\small DB, prior}}    
            \label{fig:lambdaDBprior}
        \end{subfigure}
        \hfill
        \begin{subfigure}[b]{0.495\textwidth}  
            \centering 
            \includegraphics[width=\textwidth]{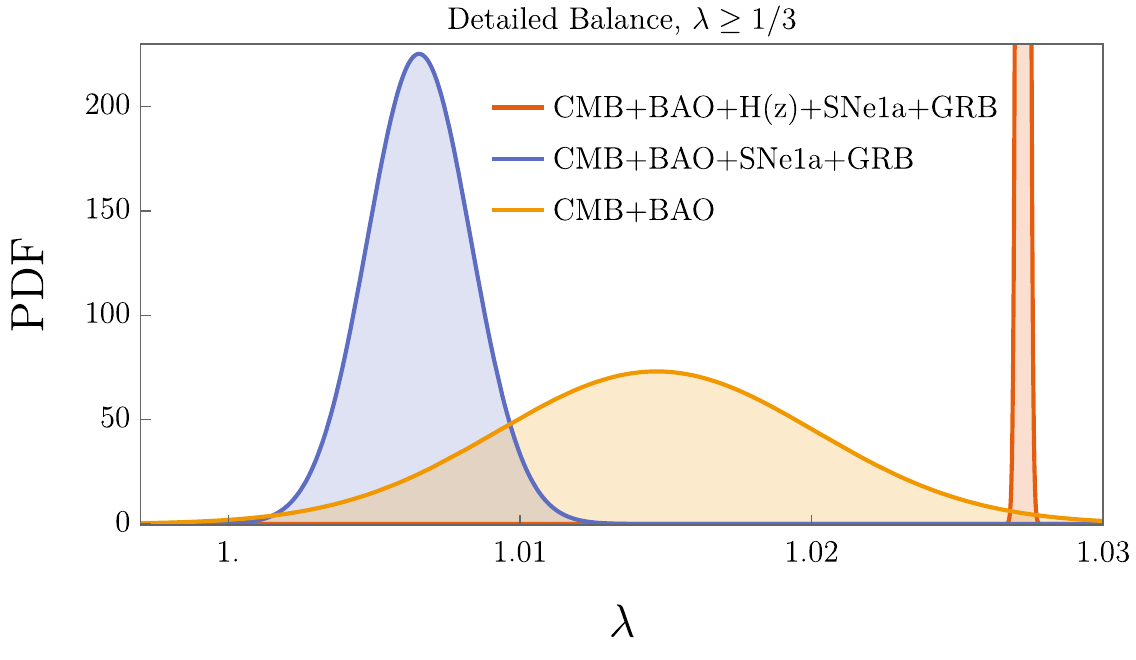}
            \caption[DB, no prior]%
            {{\small DB, no prior}}    
            \label{fig:lambdaDBnoprior}
        \end{subfigure}
        \vskip\baselineskip
        \begin{subfigure}[b]{0.495\textwidth}   
            \centering 
            \includegraphics[width=\textwidth]{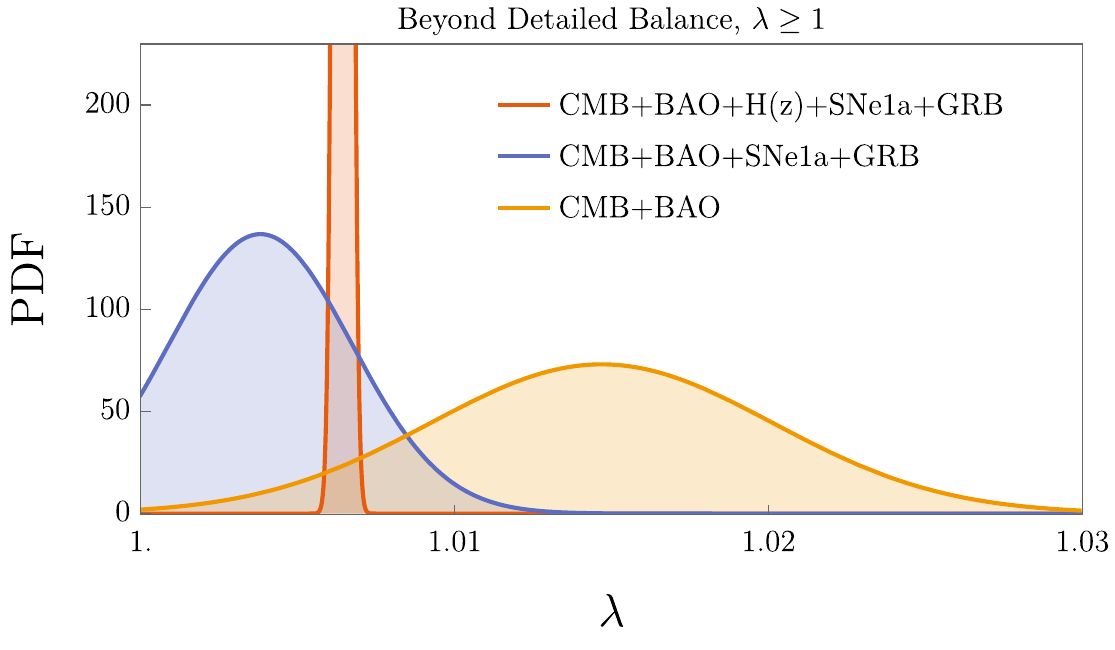}
            \caption[BDB, prior]%
            {{\small BDB, prior}}    
            \label{fig:lambdaBDBprior}
        \end{subfigure}
        \hfill
        \begin{subfigure}[b]{0.495\textwidth}   
            \centering 
            \includegraphics[width=\textwidth]{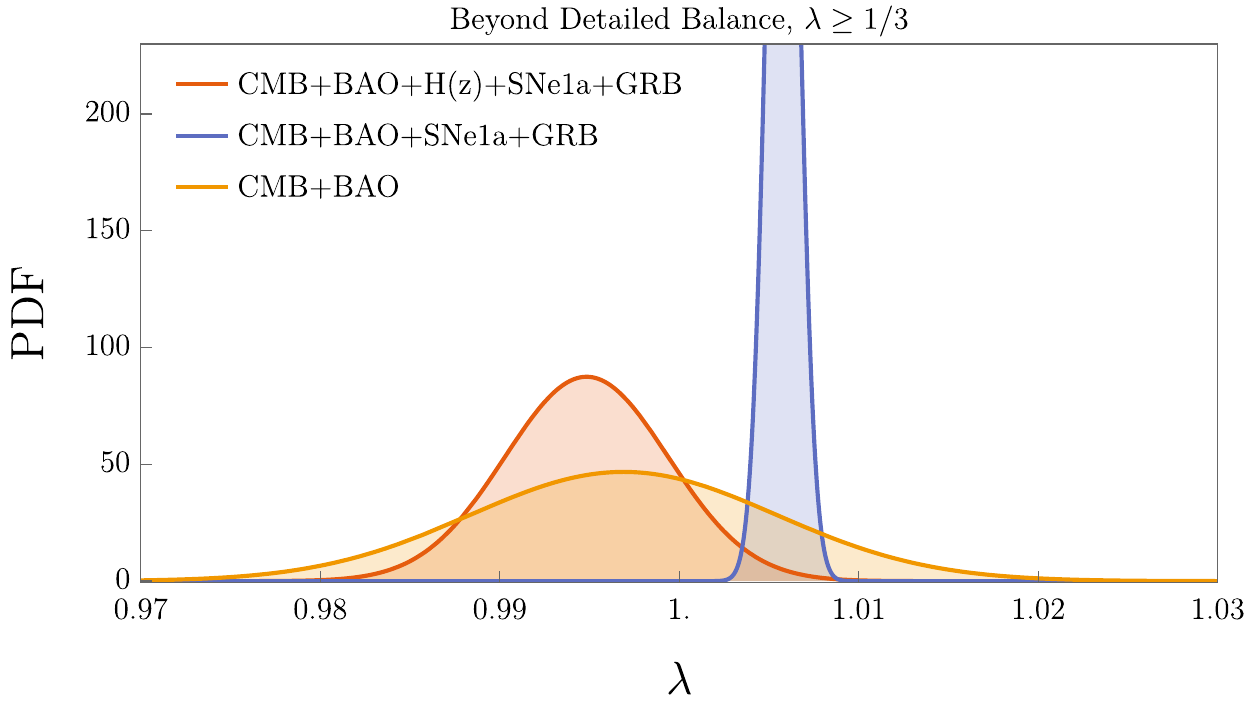}
            \caption[BDB, no prior]%
            {{\small BDB, no prior}}    
            \label{fig:lambdaBDBnoprior}
        \end{subfigure}
        \caption[]
        {\small Normalized posterior distribution functions for $\lambda$. Here, SNe1a includes Cepheid-calibrated supernovae.} 
        \label{fig:lambdas}
    \end{figure}

  {Another interesting result is the value of $\Delta N_{\rm eff}$, which takes on values greater than $0.2$ in almost all cases, and a $1\sigma$ upper limit of $\Delta N_{\rm eff}\leq0.75$ for BDB (prior, all data). As this is slightly surprising it is worth reiterating that this chain passed all convergence criteria detailed in~\cite{Dunkley:2004sv}. Therefore, with all other parameters taking on reasonable values (and similar results found in \cite{Nilsson:2018knn,Nilsson:2021ute}) one may have to entertain the possibility of a fourth neutrino species present in HL cosmology. Indeed, it was recently suggested that a fourth neutrino might solve the $H_0$
tension problem~\cite{Carneiro:2018xwq}; in this paper, the authors arrive at a value of $N_\text{eff}\approx 4$ (effective number of neutrino species), which is far higher than our results. Moreover, our results fall outside BBN limits as reported in~\cite{Hagiwara:2002fs,Steigman:2005uz} ($-1.7\leq\Delta N_{\rm eff}\leq 2.0$), but seem to agree somewhat with limits from CMB ($\Delta N_{\rm eff}<0.2$)~\cite{Cyburt:2015mya}. The fact that $\Delta N_{\rm eff}\neq 0$ in both
scenarios fits with the non-flatness results indicated by $\Omega_{k0}$. In fact, a closed Universe $\Omega_{k0}$ has been found in several different analyses of 
Ho\v{r}ava-Lifshitz cosmology \cite{Nilsson:2018knn,Nilsson:2021ute,DiValentino:2022eot}, and it now seems that the model does indeed prefer a closed Universe. Other studies have reported a strong preference for a closed Universe in the {\it Planck} data \cite{DiValentino:2019qzk}, a result which is sensitive to the amount of lensing in the sample. Other data sets, primarily BAO, strongly favour a closed Universe (under the assumption of $\Lambda$CDM) to the extent that the tension in $\Omega_{k0}$ has been estimated at $2.5-3\sigma$ \cite{Handley:2019tkm}.}

\section{Discussion \& Conclusions}

The focus of this paper is exploring the significance of the parameter $\lambda$ in HL cosmology, which is believed to play a crucial role in determining or characterizing the extent of Lorentz violation present in the theory. Our primary objective was to examine the impact of incorporating or omitting data sources corresponding to different energy levels on the estimated value of $\lambda$. In other words, we wanted to investigate whether there is a correlation between the energy level and the value of $\lambda$.

  {In the detailed-balance scenario, our findings indicate that the parameter $\lambda$ consistently exceeds unity at a 1$\sigma$ level; however, if we exclude low-energy sources, such as Hubble parameter measurements, the value of $\lambda$ tends to shift closer to the ultraviolet region. In the case of beyond detailed balance, we observed a mean value below unity in two instances. While the relationship between energy flow and $\lambda$ is not as evident here as in the detailed-balance scenario, we can still observe a clear trend: when the lowest-energy data is removed, $\lambda$ is pushed below unity.}
{Moreover, we have performed the same calculations when considering the hard prior $\lambda\ge1$,  {since analytical considerations like \cite{Bogdanos:2009uj,Horava:2009uw} demonstrate that} the parameter range $1>\lambda>1/3$ is very problematic and results in  ghost instabilities of an also problematic  scalar mode. For that reason that range is  therefore considered unphysical, however there is not yet an efficient mechanism proposed for getting rid of that problematic mode in the GR limit, which due to its higher symmetry contains only tensor modes. There are also strong observational bounds on the existence of the HL scalar mode \cite{LIGOScientific:2017ous,LIGOScientific:2018czr,Gong:2021jgg}; therefore, potential quantum instabilities of a non-observed mode and no mechanism available for suppressing it in the IR requires careful consideration before drawing any conclusions. That is why we present general results both with and without that prior on $\lambda$ in order to have a more complete picture of the theory. In the  case when the prior $\lambda\ge1$ is imposed, excluding low energy data changes the BDB case more, since this model has a stronger tendency to flow to $\lambda<1$ without a prior, but also the DB case sees changes, especially on the form of larger errors on the case with all data, although the reason for this is not clear.}

 {The phenomenological results obtained suggest the potential existence of Lorentz violation in astrophysical probes at higher energies; however, it is important to note that the existing bounds on Lorentz violation in the matter and electromagnetic sectors are significantly stronger than those in the gravity sector \cite{PhysRevLett.109.151602,Kostelecky:2008ts,Kostelecky:2010ze}. Several theoretical studies have been dedicated to addressing this issue \cite{PhysRevD.91.044021,PhysRevD.92.064037,PhysRevD.94.084014,PhysRevD.94.104043}, proposing additional terms to the action aimed at preventing the leakage of symmetry violations from the gravity sector to the matter sector or suppressing them in the low-energy regime of the matter sector.}   Moreover, there are theoretical perspectives \cite{daSilva:2010bm,Bellorin:2010je,Loll:2014xja} which question whether non-unity values of $\lambda$ necessarily indicate departures from classical general relativity. Therefore, it is crucial to exercise caution when interpreting the theoretical implications of the results presented above. 
 A natural question which also arises is how the parameter $\lambda$ would vary in different non-projectable extensions of HL gravity, which have been proposed to address some of its shortcomings. While work on these scenarios is ongoing, it is important to note that relaxing projectability conditions leads to a more complex theory with a higher number of parameters and interpretation issues. For now, we present our current findings, which are intriguing and merit further theoretical and observational investigation. These results could offer new insights into the underlying HL theory.

\begin{acknowledgements}
NAN is thankful for financial support by CNES, and for discussions with Vincenzo Salzano.
\end{acknowledgements}

\appendix

\section{The data}
\label{app:data}
\subsection{{\it Planck} CMB}
  {The non-perturbative CMB information is contained in the shift parameters, describing the location of the first peak in the temperature angular power spectrum. Here, we use the shift parameters extracted from the final {\it Planck} 2018 data release \cite{Planck:2018vyg}. For a vector $\bm{\theta}$ containing the model parameters, the geometrical CMB shift parameters read
\begin{equation}
    R(\bm{\theta})=100\sqrt{\Omega_{m0}h^2}\frac{d_A^c(z_*,\bm{\theta})}{c}, \quad \ell_a(\bm{\theta}) = \pi \frac{d_A^c(z_*,\bm{\theta})}{r_s(z_*,\bm{\theta})},
\end{equation}
which together with $\omega_b = \Omega_{b0}h^2$ makes up the CMB distance priors. In the above equations, $d_A^c(z_*,\bm{\theta})$ and $r_s(z_*,\bm{\theta})$ are the comoving angular-diameter distance and the sound horizon, respectively, defined as
\begin{equation}
    d_A^c(z,\bm{\theta}) =
    \begin{cases}
    \frac{c}{H_0}\frac{1}{\sqrt{\Omega_{k0}}}\sinh{\left[\sqrt{\Omega_{k0}}\int_0^z\frac{dz'}{E(z',\bm{\theta})}\right]}, &\Omega_{k0}>0 \\
    \frac{c}{H_0} \int_0^z\frac{dz'}{E(z',\bm{\theta})}, &\Omega_{k0}=0 \\
    \frac{c}{H_0}\frac{1}{\sqrt{|\Omega_{k0}|}}\sinh{\left[\sqrt{|\Omega_{k0}|}\int_0^z\frac{dz'}{E(z',\bm{\theta})}\right]}, &\Omega_{k0}>0,
    \end{cases}
\end{equation}
and
\begin{equation}
    r_s(z,\bm{\theta}) = H_0\int_z^\infty \frac{c_s(z')dz'}{E(z',\bm{\theta})},
\end{equation}
where $E(z,\bm{\theta})\equiv H(z,\bm{\theta})/H_0$ and $c_s(z)$ is the sound speed
\begin{equation}
    c_s(z) = \frac{c}{\sqrt{3[1+R_b/(1+z)]}}, \quad R_b=31500\omega_b\left(\frac{T_{\rm CMB}}{2.72}\right)^{-4}.
\end{equation}
The above relations are defined at the photon-decoupling redshift $z_*$, which is defined as \cite{Hu:1995en}
\begin{equation}
    \begin{aligned}
        z_* =& 1048\left[1+0.00124\omega_b^{-0.738}\right]\left[1+g_1 \omega_b^{g_2}\right] \\
        g_1 =& 0.0783 \omega_b^{-0.238}\left[1+39.5 \omega_b^{-0.763}\right]^{-1} \\
        g_2 =& 0.560\left[1+21.1\omega_b^{1.81}\right]^{-1}.
    \end{aligned}
\end{equation}
}

  {In the model we are considering, the number of relativistic species are no longer equal to the $\Lambda$CDM value of $N_{\rm eff}=3.046$, and since the value of this quantity have several effects on the CMB, we need now to incorporate this into the analysis. Following the procedure in \cite{Zhai:2019nad}, we add the CDM density parameter $\omega_c = (\Omega_{m0}-\Omega_{b0})h^2$ and $N_{\rm eff} = 3.046+\Delta N_{\rm eff}$, and the full CMB distance priors read $\bm{v} = (R,\ell_a,\omega_b,\omega_c,N_{\rm eff})$. We use the values found in \cite{Zhai:2019nad}
\begin{equation}
    \bm{v} = \begin{pmatrix}
    1.7661\\301.7293\\0.02191\\0.1194\\2.8979,
    \end{pmatrix}
\end{equation}
as well as the associated covariance matrix
\begin{equation}
    C_{\bm{v}} = 10^{-8}\begin{pmatrix}
        33483.54 & -44417.15 & -515.03 & -360.42 & -274151.72 \\
        -44417.15 & 4245661.67 & 2319.46 & 63326.47 & 4287810.44 \\
        -515.03 & 2319.46 & 12.92 & 51.98 & 7273.04 \\
        -360.42 & 63326.47 & 51.98 & 1516.28 & 92013.95 \\
        -274151.72 & 4287810.44 & 7273.04 & 92013.95 & 7876074.60
    \end{pmatrix}
\end{equation}
and the CMB $\chi^2$ finally reads
\begin{equation}
    \chi^2_{\rm CMB} = \left(\Delta\bm{v}\right)^{\rm T} C_{\bm{v}}^{-1}\Delta\bm{v},
\end{equation}
where $\Delta\bm{v}$ is the difference between the theoretical and observed values of the distance priors $\bm{v}$.}

\subsection{Pantheon+ and Cepheid variable stars}
  {The Pantheon+ sample is a set of 1701 light curves of 1550 distinct Supernovae Type Ia (SNeIa) in the redshift range $0.001<z<2.26$ \cite{Scolnic:2021amr,Brout:2022vxf}. Compared to the previous catalogue Pantheon, this update features an increased redshift range of cross-calibrated photometric systems of sources and improved treatment of systematic effects; all together, this results in an factor of 2 improvement in cosmological constraining power \cite{Brout:2022vxf}. Also contained in this catalogue are those SNeIa from host galaxies with known Cepheid distances, providing a robust calibration of the SNeIa light curve (known as ``anchoring'') and enables the simultaneous determination of expansion-history parameters (for example $\Omega_{m0}$) and the local expansion rate $H_0$, which are degenerate when using SNeIa alone. }

  {In order to form the $\chi^2$ for Pantheon, we write down the luminosity distance as (all details can be found in \cite{Brout:2022vxf})
\begin{equation}
    d_L(z,\bm{\theta})=(1+z_{\rm hel})\,d_A^c(z,\bm{\theta}),
\end{equation}
where $z_{\rm hel}$ is the redshift measured in the heliocentric frame. From this we can define distance modulus as
\begin{equation}
    \mu(z,\bm{\theta}) = m(z,\bm{\theta})-M,
\end{equation}
where $m$ and $M$ is the apparent and fiducial absolute magnitude, respectively.
From this expression, we can write the apparent magnitude as
\begin{equation}
    m(z,\bm{\theta}) = 5 \log{d_L(z,\bm{\theta})}+25+M,
\end{equation}
where $d_L(z,\bm{\theta})$ is now expressed in Mpc. Note here that we are not able to simply write the distance modulus as $\mu=m-M = 5\log{d_L} +\mu_0$ and marginalize over $\mu_0$; when using SNeIa alone, $M$ can indeed be marginalized over, but including Cepheids breaks the degeneracy between $M$ and $H_0$, and the fiducial absolute magnitude $M$ becomes an extra free parameter in our numerical analysis. As such, we write the distance residuals as \cite{Brout:2022vxf}
\begin{equation}
    \Delta D_i = \begin{cases}
        \mu_i - \mu_i^C, & i\in \text{Cepheids} \\
        \mu_i - \mu_i^{\rm model} & \text{others},
    \end{cases}
\end{equation}
where the theoretical value is replaced by the corresponding Cepheid-calibrated value for the host galaxy distance, hence providing the anchoring and breaking the degeneracy between $\Omega_{m0}$ and $M$. The statistical and systematic uncertainties are contained in the covariance matrix $C^{\rm SN+Cepheids}_{\rm stat+sys}$, and the $\chi^2$ measure becomes
\begin{equation}
    \chi^2_{\rm Pantheon+Cepheids} = (\Delta \bm{D})^{\rm T} \left(C^{\rm SN+Cepheids}_{\rm stat+sys}\right)^{-1}(\Delta \bm{D})
\end{equation}
}

\subsection{Gamma-ray bursts}
  {Gamma-ray bursts are one of the most energetic events in the Universe, and may be used as a probe complementing SNeIa~\cite{Liu:2014vda}. This sample consists of 79 long gamma-ray bursts (the Mayflower sample) calibrated using the Padé approximant method. This approach avoids the circularity problem usually present when trying to use gamma-ray bursts as cosmological rulers~\cite{Ghirlanda:2006ax}. The $\chi^2$ for this dataset is found by considering the distance modulus (as for SNeIa, but we can now marginalize over $\mu_0$), and we write the $\chi^2$ as
\begin{equation}
    \chi^2_{\rm GRB} = a+\log{\frac{e}{2\pi}}-\frac{b^2}{e},
\end{equation}
where $a=(\Delta\mu)^{\rm T} C^{-1}_{\rm GRB} (\Delta\mu)$, $b = (\Delta\mu)^{\rm T}C^{-1}_{\rm GRB} \cdot \bm{1}$, and $c = \bm{1}^{\rm T}\cdot C^{-1}_{\rm GRB}\cdot \bm{1}$. Here we have defined $\Delta\mu = \mu_{\rm theory}-\mu_{\rm obs}$
where $\mu_{\rm obs}$ is the observed distance modulus, and $\mu_{\rm theory}$ comes from the Padé method after calibration. The full details of this calibration and the data can be found in~\cite{Liu:2014vda}.}

\subsection{Cosmic Chronometers (CC)}
  {We use Hubble-parameter measurements from Passively-evolving Early-Type Galaxies (ETG), which have low star-formation rate and old stellar populations. The spectral properties of ETGs can be traced along cosmic time $t$ by measuring the Hubble parameter $H(z) = dz/dt(1+z)$ independent of the cosmological model~\cite{Jimenez:2001gg}, making them a type of standardisable clock, or cosmic chronometer. We use a sample covering the redshift range $0<z<1.97$ \cite{Moresco:2012by,Moresco:2015cya,Gomez-Valent:2019lny}. To construct a covariance matrix for the CC data points we follow the procedure in \cite{Moresco:2022phi,Moresco:2020fbm}, which includes the following sources of uncertainty
\begin{equation}
        C_{\rm CC} = C_{\rm CC}^{\rm stat}+C_{\rm CC}^{\rm young}+C_{\rm CC}^{\rm model}+C_{\rm CC}^{\rm met},
\end{equation}
where $C_{\rm CC}^{\rm stat}$, $C_{\rm CC}^{\rm young}$, and $C_{\rm CC}^{\rm model}$ correspond to uncertainty from statistical errors, sample-contamination by younger (and hotter) stars, model dependence, and stellar metallicity. In these considerations, the largest source of error comes from the $C_{\rm CC}^{\rm young}$ contribution \cite{Moresco:2020fbm}. 
The model uncertainty $C_{\rm CC}^{\rm model}$ can be further broken down into
\begin{equation}
    C_{\rm CC}^{\rm model} = C_{\rm CC}^{\rm SFH}+C_{\rm CC}^{\rm IMF}+C_{\rm CC}^{\rm st.lib.}+C_{\rm CC}^{\rm SPS},
\end{equation}
denoting star-formation history (SFH), initial mass function (IMF), stellar library (st.lib.), and stellar population synthesis (SPS). We use the accompanying code\footnote{\url{https://gitlab.com/mmoresco/CCcovariance}} \cite{Moresco:2020fbm} to generate the final covariance matrix for our data points, after which the $\chi^2$ reads
\begin{equation}
    \chi^2_{\rm CC} = (\Delta H)^{\rm T} \, C_{\rm CC}^{-1} \, (\Delta H),
\end{equation}
where $(\Delta H)_i = H_{\rm theory}(z_i)-H_{\rm obs}(z_i)$.
}

\subsection{Baryon Acoustic Oscillations}
  {We include seven different BAO data sets in our analysis, as shown below}

  {\textbf{WiggleZ: } We include data from the WiggleZ Dark Energy Survey at redshift points $z_W=\{0.44,0.6,0.73\}$ \cite{2012MNRAS.425..405B}. Here, the observables are
\begin{equation}
    A(x,\bm{\theta})=100\sqrt{\omega_m}\frac{D_V(z,\bm{\theta})}{cz}, \quad F(z,\bm{\theta})=\frac{d_A^c(z,\bm{\theta})H(z,\bm{\theta})}{c},
\end{equation}
where $A(x,\bm{\theta})$ is the acoustic parameter, $F(z,\bm{\theta})$ is the Alcock-Paczynski distortion parameter, and $\omega_m=\Omega_{m0}h^2$. $D_V(z,\bm{\theta})$ is the volume distance, defined as
\begin{equation}
    D_V(z,\bm{\theta})=\left[d_A^c(z,\bm{\theta})^2\frac{cz}{H(z,\bm{\theta})}\right]^{1/3},
\end{equation}
and we find the $\chi^2$ as
\begin{equation}
    \chi^2_W = (\Delta\mathcal{F}_W)^{\rm T} C^{-1}_W \Delta\mathcal{F}_W,
\end{equation}
where $\Delta\mathcal{F}_W = \mathcal{F}_{W, \rm theory}-\mathcal{F}_{W, \rm obs}$, and $\mathcal{F}_W = \{A(z_W), F(z_W)\}$. For all the other BAO probes, we find the $\chi^2$ in the same way.}

  {\textbf{SDSS-BOSS: } we include data from the SDSS-II BOSS DR12 and SDSS-IV DR16 LRG (Luminous Red Galaxy growth-rate sample) at redshifts $z_{\rm B}=\{0.38,0.51,0.61\}$. For this data, we use the quantities
\begin{equation}\label{eq:dah}
    d_A^c(z,\bm{\theta})\frac{r_s^{\rm fid}(z_d)}{r_s(z_d,\bm{\theta})}, \quad H(z,\bm{\theta})\frac{r_s(z_d,\bm{\theta})}{r _s^{\rm fid}(z_d)},
\end{equation}
evaluated at the dragging redshift $z_d$, which we approximate as \cite{Eisenstein:1997ik}
\begin{equation}
    \begin{aligned}
        z_d =& 1291 \frac{\omega_m^{~0.251}}{1+0.659 \omega_m^{~0.828}}\left(1+b_1 \omega_b^{~b_2}\right) \\
        b_1 =& 0.313 \omega_m^{~-0.419}\left(1+0.607\omega_m^{~0.6748}\right) \\
        b_2 =& 0.238 \omega_m^{~0.223}.
    \end{aligned}
\end{equation}
In Eq.(\ref{eq:dah}) $r_s^{\rm fid}$ is the sound horizon at the dragging redshift evaluated for a fiducial cosmological model; here, we take $r_s^{\rm fid}(z_d)=147.78$ Mpc \cite{Bautista:2020ahg,BOSS:2016wmc}.}

  {\textbf{SDSS QSO: } From the SDSS IV BOSS DR14 quasar sample, we have the following data points at redshifts $z_Q=\{0.978,1.23,1.526,1.944\}$ \cite{Zhao:2018gvb}
\begin{equation}
\begin{aligned}
    d_A^c\frac{r_s^{\rm fid}(z_d)}{r_s(z_d} =& \{1586.18\pm 284.93, 1769.08 \pm 159.67, 1768.77 \pm 96.59, 1807.98 \pm 146.46\} \\
    H\frac{r_s(z_d)}{r _s^{\rm fid}(z_d)} =& \{113.72 \pm 14.63, 131.44 \pm 12.42, 148.11 \pm 12.75, 172.63 \pm 14.79\} \\
    D_V\frac{r_s^{\rm fid}(z_d)}{r_s(z_d)} =& \{2933.59 \pm 327.71, 3522.04 \pm 192.74, 3954.31 \pm 141.71, 4575.17 \pm 241.61\}.
\end{aligned}
\end{equation}
From the eBOSS DR16 QSO release \cite{Hou:2020rse,Neveux:2020voa}, we also have the following data points at redshift $z=1.480$
\begin{equation}
    \frac{c}{H r_s(z_d)} = 13.23\pm 0.47, \quad \frac{d_A^c}{r_s(z_d)} = 30.21\pm0.79.
\end{equation}}

  {\textbf{eBOSS ELG: } from the eBOSS DR16 Emission Line Galaxy sample (ELG) we have, at the effective redshift $z_{\rm eff}=0.845$ \cite{Tamone:2020qrl,deMattia:2020fkb}
\begin{equation}
    \frac{c}{H r_s(z_{\rm eff})} = 19.6^{+2.2}_{-2.1}, \quad \frac{d_A^c(z_{\rm eff})}{r_s(z_d)} = 19.5\pm 1
\end{equation}}

  {\textbf{eBOSS CMASS: } from void-galaxy cross-correlations in redshift-space distortion corrected data from the DR12 LRG CMASS sample, we have, at the effective sample redshift of $z_{\rm eff}=0.69$ the following \cite{Nadathur:2020vld}
\begin{equation}
    \frac{c}{H r_s(z_{\rm eff})} = 17.48\pm0.23, \quad \frac{d_A^c(z_{\rm eff})}{r_s(z_d)} = 20.10\pm 0.34.
\end{equation}}

  {\textbf{Lyman-$\alpha$: } using the autocorrelation of Lyman-$\alpha$ absorption in quasars and Lyman-$\alpha$ cross correlation in the eBOSS DR16 quasar sample \cite{duMasdesBourboux:2020pck} we have (at the effective redshift $z_{\rm eff}=2.33$)
\begin{equation}
    \frac{c}{H r_s(z_{\rm eff})} = 8.99\pm0.19, \quad \frac{d_A^c(z_{\rm eff})}{r_s(z_d)} = 37.5\pm 1.1.
\end{equation}}

  {All the $\chi^2$ measures for the above BAO points are included in our analysis.}
\clearpage
 \section{All parameter constraints}\label{app:tables}
 
 \begingroup
\squeezetable
\begin{center}
\begin{table*}[h]
\renewcommand{\arraystretch}{2}
%\begin{tabular}{l|@{\hspace{0.5 cm}} ccc | ccc}
\begin{tabular}{l|@{\hspace{0.5 cm}} ccc @{\hspace{3mm}}|@{\hspace{1mm}}ccc}
\multicolumn{1}{c}{} & \multicolumn{3}{c}{Detailed balance, prior $\lambda\geq1$} & \multicolumn{3}{c}{Detailed balance, no prior on $\lambda$} \\
\hline\hline
Parameter & \makecell{CMB+BAO+H(z)\\+SNe1a+Cepheids\\+GRB} & \makecell{CMB+BAO\\+SNe1a+Cepheids\\+GRB} & \makecell{CMB+BAO} & \makecell{CMB+BAO+H(z)\\+SNe1a+Cepheids\\+GRB} & \makecell{CMB+BAO\\+SNe1a+Cepheids\\+GRB} & \makecell{CMB+BAO} \\
\hline
$\Omega_b$ \dotfill & $0.0049768\pm0.0000016$ & $0.04898^{+0.00055}_{-0.000053}$ & $0.05104\pm0.00080$ & $0.049236\pm0.000070$ & $0.04907\pm0.00058$ & $0.05139^{+0.00019}_{-0.00020}$ \\
$\Omega_b h^2$ \dotfill & $0.020952\pm0.000074$ & $0.02227\pm0.00018$ & $0.02192^{+0.00019}_{-0.00021}$ & $0.02158\pm0.00012$ & $0.02217\pm0.00017$ & $0.02187\pm0.00011$ \\
$\Omega_m$ \dotfill & $0.3336\pm0.0018$ & $0.3153^{+0.0057}_{-0.0055}$ & $0.3398\pm0.0083$ & $0.3204\pm0.0030$ & $0.3170\pm0.0056$ & $0.34437^{+0.00010}_{-0.00012}$ \\
$\Omega_m h^2$ \dotfill & $0.14043^{+0.00024}_{-0.00025}$ & $0.14335\pm0.00090$ & $0.1458^{+0.0011}_{-0.0010}$ & $0.14047^{+0.00030}_{-0.00033}$ & $0.14323\pm0.00092$ & $0.14657\pm0.00023$ \\
$\Omega_k 10^{4}$ \dotfill & $-4.1364\pm0.0040$ & $-6.116^{+1.55}_{-0.30}$ & $-11.60^{+2.61}_{-1.76}$ & $-4.254\pm0.019$ & $-5.745\pm0.029$ & $-13.5030^{+0.0088}_{-0.0086}$ \\
$\Omega_r 10^{5}$ \dotfill & $9.937^{+0.036}_{-0.035}$ & $9.20\pm0.12$ & $9.74^{+0.19}_{-0.18}$ & $9.543^{+0.070}_{-0.065}$ & $9.259^{+0.012}_{-0.011}$ & $9.83^{+0.012}_{-0.018}$ \\
$h$ \dotfill & $0.6488\pm0.0012$ & $0.6743^{+0.0043}_{-0.0044}$ & $0.6553^{+0.0061}_{-0.0063}$ & $0.6621^{+0.0023}_{-0.0024}$ & $0.6722\pm0.0040$ & $0.65239\pm0.00040$ \\
$M$ \dotfill & $-19.5051\pm0.0013$ & $-19.437^{+0.012}_{-0.013}$ & - & $-19.4783\pm0.0075$ & $-19.442\pm0.012$ & - \\
$\Delta N_{\rm eff}$ \dotfill & $0.0046750\pm0.0000076$ & $\bm{0.1104^{+0.0011}_{-0.0049}}$ & $0.038^{+0.013}_{-0.015}$ & $0.005099\pm0.000060$ & $0.009670\pm0.000062$ & $0.05195\pm0.00019$ \\
$\lambda$ \dotfill & $1.0406\pm0.0023$ & $\bm{<1.0032}^\dagger$ & $1.0146^{+0.055}_{-0.053}$ & $1.02726\pm0.00012$ & $1.0065\pm0.0018$ & $1.0159\pm0.0014$ \\
$\left|\tfrac{G_{\rm cosmo}}{G_{\rm grav}}-1\right|$ \dotfill & $0.0574\pm0.0030$ & $\bm{<0.0035}^\dagger$ & $0.0214^{+0.0078}_{-0.0077}$ & $0.03928\pm0.00017$ & $0.00997^{+0.0026}_{-0.0025}$ & $0.0232\pm0.0020$ \\
\hline
\hline
$\chi^2_{\rm{min}}$ \dotfill & $1778.27$ & $1635.41$ & $27.30$ & $1705.04$ & $1638.49$  & $27.76$  \\
\hline\hline
\end{tabular}
\caption{Parameter constraints at $1\sigma$ for the Detailed Balance case, with and without a hard prior on the parameter $\lambda$. $\dagger$ implies a one-sided upper bound resulting from a hard uniform prior, and \textbf{bold} indicates a particularly noisy parameter.
}
\label{tab:DB}
\end{table*}
\end{center}
\endgroup

 \begingroup
\squeezetable
\begin{center}
\begin{table*}[h]
\renewcommand{\arraystretch}{2}
\begin{tabular}{l|@{\hspace{0.5 cm}} ccc @{\hspace{3mm}}|@{\hspace{1mm}}ccc}
\multicolumn{1}{c}{} & \multicolumn{3}{c}{Beyond detailed balance, prior $\lambda\geq1$} & \multicolumn{3}{c}{Beyond detailed balance, no prior on $\lambda$} \\
\hline\hline
Parameter & \makecell{CMB+BAO+H(z)\\+SNe1a+Cepheids\\+GRB} & \makecell{CMB+BAO\\+SNe1a+Cepheids\\+GRB} & \makecell{CMB+BAO} & \makecell{CMB+BAO+H(z)\\+SNe1a+Cepheids\\+GRB} & \makecell{CMB+BAO\\+SNe1a+Cepheids\\+GRB} & \makecell{CMB+BAO} \\
\hline
$\Omega_b$ \dotfill & $0.049371^{+0.00049}_{-0.00048}$ & $0.05024\pm0.00049$ & $0.05116^{+0.00062}_{-0.00063}$ & $0.04900\pm0.00040$ & $0.05034^{+0.00017}_{-0.00016}$ & $0.05034\pm0.00016$ \\
$\Omega_b h^2$ \dotfill & $0.022922\pm0.00023$ & $0.02262\pm0.00020$ & $0.02226\pm0.00018$ & $0.022864\pm0.00011$ & $0.02250\pm0.00011$ & $0.02250\pm0.00011$ \\
$\Omega_m$ \dotfill & $0.3198\pm0.0053$ & $0.3194^{+0.0056}_{-0.0054}$ & $0.3349\pm0.0073$ & $0.3109^{+0.0038}_{-0.0041}$ & $0.3232\pm0.0031$ & $0.3284^{+0.0093}_{-0.0091}$ \\
$\Omega_m h^2$ \dotfill & $0.1484\pm0.0017$ & $0.14386^{+0.00095}_{-0.00096}$ & $0.1458\pm0.0011$ & $0.14509^{+0.00015}_{-0.00016}$ & $0.14442^{+0.00046}_{-0.00049}$ & $0.1458^{+0.0030}_{-0.0027}$ \\
$\Omega_k 10^{3}$ \dotfill & $-9.71^{+1.73}_{-1.83}$ & $-4.98^{+0.77}_{-0.46}$ & $-3.93\pm0.15$ & $-5.399^{+0.022}_{-0.023}$ & $-4.338\pm0.080$ & $-6.32\pm2.56$ \\
$\Omega_r 10^{5}$ \dotfill & $9.01\pm0.13$ & $9.29\pm0.11$ & $9.61^{+1.63}_{-1.61}$ & $8.97^{+1.06}_{-1.11}$ &  $9.36\pm0.61$ & $9.431^{+0.028}_{-0.030}$ \\
$h$ \dotfill & $0.6813\pm0.0048$ & $0.6711^{+0.0039}_{-0.0040}$ & $0.6599^{+0.0056}_{-0.0055}$ & $0.6831^{+0.0043}_{-0.0040}$ & $0.6685^{+0.0022}_{-0.0021}$ & $0.6660^{+0.0109}_{-0.0096}$ \\
$M$ \dotfill & $-19.414^{+0.014}_{-0.015}$ & $-19.446\pm0.0011$ & - & $-19.412^{+0.012}_{-0.011}$ & $-19.4525\pm0.0040$ & - \\
$\Delta N_{\rm eff}$ \dotfill & $0.61\pm0.14$ & $0.258^{+0.028}_{-0.048}$ & $0.198\pm0.015$ & $0.2578^{+0.0042}_{-0.0044}$ & $0.2166\pm0.0070$ & $0.31^{+0.19}_{-0.15}$ \\
$\lambda$ \dotfill  & $1.00644\pm0.00020$ & $<1.0068^\dagger$ & $1.0065\pm0.0025$ & $0.9949^{+0.0045}_{-0.0046}$ & $1.00578\pm0.00086$ & $0.9972^{+0.0081}_{-0.0088}$ \\
$\left|\tfrac{G_{\rm cosmo}}{G_{\rm grav}}-1\right|$ \dotfill & $0.00957\pm0.00029$ & $<0.010^\dagger$ & $0.0096\pm0.0037$ & $0.0078^{+0.0070}_{-0.0053}$ & $0.0086\pm0.0013$ & $<0.019$ \\
$\omega_1$ \dotfill & $0.69957\pm0.00039$ & $0.6909^{+0.0063}_{-0.0062}$ & $0.6789^{+0.0070}_{-0.0075}$ & $0.6866\pm0.0036$ & $0.6898^{+0.0044}_{-0.0043}$ & $0.6898\pm0.0043$ \\
$\omega_3 10^{6}$ \dotfill & $7.07^{+1.62}_{-1.71}$ & $1.77^{+0.10}_{-0.15}$ & $1.5201^{+0.068}_{-0.071}$ & $2.9109^{+0.047}_{-0.0049}$ & $1.80\pm0.15$ & $1.79\pm0.15$ \\
\hline
\hline
$\chi^2_{\rm{min}}$ \dotfill & $1634.37$ & $1632.36$ & $23.82$ & $1635.54$ & $1633.85$ & $21.85$ \\
\hline\hline
\end{tabular}
\caption{Parameter constraints at $1\sigma$ for the Beyond Detailed Balance case, with and without a hard prior on the parameter $\lambda$. $\dagger$ implies a one-sided upper bound resulting from a hard uniform prior.
}
\label{tab:BDB}
\end{table*}
\end{center}
\endgroup

\bibliography{horavalambda.bib}

%apsrev4-2.bst 2015-08-30 from 4.21a (PWD, AO, DPC/HNN) hacked
\begin{thebibliography}{99}%
\makeatletter
\providecommand \@ifxundefined [1]{%
 \@ifx{#1\undefined}
}%
\providecommand \@ifnum [1]{%
 \ifnum #1\expandafter \@firstoftwo
 \else \expandafter \@secondoftwo
 \fi
}%
\providecommand \@ifx [1]{%
 \ifx #1\expandafter \@firstoftwo
 \else \expandafter \@secondoftwo
 \fi
}%
\providecommand \natexlab [1]{#1}%
\providecommand \enquote  [1]{``#1''}%
\providecommand \bibnamefont  [1]{#1}%
\providecommand \bibfnamefont [1]{#1}%
\providecommand \citenamefont [1]{#1}%
\providecommand \href@noop [0]{\@secondoftwo}%
\providecommand \href [0]{\begingroup \@sanitize@url \@href}%
\providecommand \@href[1]{\@@startlink{#1}\@@href}%
\providecommand \@@href[1]{\endgroup#1\@@endlink}%
\providecommand \@sanitize@url [0]{\catcode `\\12\catcode `\$12\catcode
  `\&12\catcode `\#12\catcode `\^12\catcode `\_12\catcode `\%12\relax}%
\providecommand \@@startlink[1]{}%
\providecommand \@@endlink[0]{}%
\providecommand \url  [0]{\begingroup\@sanitize@url \@url }%
\providecommand \@url [1]{\endgroup\@href {#1}{\urlprefix }}%
\providecommand \urlprefix  [0]{URL }%
\providecommand \Eprint [0]{\href }%
\providecommand \doibase [0]{http://dx.doi.org/}%
\providecommand \selectlanguage [0]{\@gobble}%
\providecommand \bibinfo  [0]{\@secondoftwo}%
\providecommand \bibfield  [0]{\@secondoftwo}%
\providecommand \translation [1]{[#1]}%
\providecommand \BibitemOpen [0]{}%
\providecommand \bibitemStop [0]{}%
\providecommand \bibitemNoStop [0]{.\EOS\space}%
\providecommand \EOS [0]{\spacefactor3000\relax}%
\providecommand \BibitemShut  [1]{\csname bibitem#1\endcsname}%
\let\auto@bib@innerbib\@empty
%</preamble>
\bibitem [{\citenamefont {Horava}(2009)}]{Horava:2009if}%
  \BibitemOpen
  \bibfield  {author} {\bibinfo {author} {\bibfnamefont {P.}~\bibnamefont
  {Horava}},\ }\bibfield  {title} {\enquote {\bibinfo {title} {{Spectral
  Dimension of the Universe in Quantum Gravity at a Lifshitz Point}},}\ }\href
  {\doibase 10.1103/PhysRevLett.102.161301} {\bibfield  {journal} {\bibinfo
  {journal} {Phys. Rev. Lett.}\ }\textbf {\bibinfo {volume} {102}},\ \bibinfo
  {pages} {161301} (\bibinfo {year} {2009})},\ \Eprint
  {http://arxiv.org/abs/0902.3657}{arXiv:0902.3657 [hep-th]}\BibitemShut
  {NoStop}%
%%CITATION = ARXIV:0902.3657;%%
\bibitem [{\citenamefont {Bluhm}(2015)}]{Bluhm:2014oua}%
  \BibitemOpen
  \bibfield  {author} {\bibinfo {author} {\bibfnamefont {R.}~\bibnamefont
  {Bluhm}},\ }\bibfield  {title} {\enquote {\bibinfo {title} {{Explicit versus
  Spontaneous Diffeomorphism Breaking in Gravity}},}\ }\href {\doibase
  10.1103/PhysRevD.91.065034} {\bibfield  {journal} {\bibinfo  {journal} {Phys.
  Rev. D}\ }\textbf {\bibinfo {volume} {91}},\ \bibinfo {pages} {065034}
  (\bibinfo {year} {2015})},\ \Eprint
  {http://arxiv.org/abs/1401.4515}{arXiv:1401.4515 [gr-qc]}\BibitemShut
  {NoStop}%
\bibitem [{\citenamefont {Sotiriou}\ \emph {et~al.}(2009)\citenamefont
  {Sotiriou}, \citenamefont {Visser},\ and\ \citenamefont
  {Weinfurtner}}]{Sotiriou:2009bx}%
  \BibitemOpen
  \bibfield  {author} {\bibinfo {author} {\bibfnamefont {T.~P.}\ \bibnamefont
  {Sotiriou}}, \bibinfo {author} {\bibfnamefont {M.}~\bibnamefont {Visser}}, \
  and\ \bibinfo {author} {\bibfnamefont {S.}~\bibnamefont {Weinfurtner}},\
  }\bibfield  {title} {\enquote {\bibinfo {title} {{Quantum gravity without
  Lorentz invariance}},}\ }\href {\doibase 10.1088/1126-6708/2009/10/033}
  {\bibfield  {journal} {\bibinfo  {journal} {JHEP}\ }\textbf {\bibinfo
  {volume} {10}},\ \bibinfo {pages} {033} (\bibinfo {year} {2009})},\ \Eprint
  {http://arxiv.org/abs/0905.2798}{arXiv:0905.2798 [hep-th]}\BibitemShut
  {NoStop}%
%%CITATION = ARXIV:0905.2798;%%
\bibitem [{\citenamefont {Arnowitt}\ \emph {et~al.}(2008)\citenamefont
  {Arnowitt}, \citenamefont {Deser},\ and\ \citenamefont
  {Misner}}]{Arnowitt:1962hi}%
  \BibitemOpen
  \bibfield  {author} {\bibinfo {author} {\bibfnamefont {R.~L.}\ \bibnamefont
  {Arnowitt}}, \bibinfo {author} {\bibfnamefont {S.}~\bibnamefont {Deser}}, \
  and\ \bibinfo {author} {\bibfnamefont {C.~W.}\ \bibnamefont {Misner}},\
  }\bibfield  {title} {\enquote {\bibinfo {title} {{The Dynamics of general
  relativity}},}\ }\href {\doibase 10.1007/s10714-008-0661-1} {\bibfield
  {journal} {\bibinfo  {journal} {Gen. Rel. Grav.}\ }\textbf {\bibinfo {volume}
  {40}},\ \bibinfo {pages} {1997} (\bibinfo {year} {2008})},\ \Eprint
  {http://arxiv.org/abs/gr-qc/0405109}{arXiv:gr-qc/0405109}\BibitemShut
  {NoStop}%
\bibitem [{\citenamefont {Bogdanos}\ and\ \citenamefont
  {Saridakis}(2010)}]{Bogdanos:2009uj}%
  \BibitemOpen
  \bibfield  {author} {\bibinfo {author} {\bibfnamefont {C.}~\bibnamefont
  {Bogdanos}}\ and\ \bibinfo {author} {\bibfnamefont {E.~N.}\ \bibnamefont
  {Saridakis}},\ }\bibfield  {title} {\enquote {\bibinfo {title} {{Perturbative
  instabilities in Horava gravity}},}\ }\href {\doibase
  10.1088/0264-9381/27/7/075005} {\bibfield  {journal} {\bibinfo  {journal}
  {Class. Quant. Grav.}\ }\textbf {\bibinfo {volume} {27}},\ \bibinfo {pages}
  {075005} (\bibinfo {year} {2010})},\ \Eprint
  {http://arxiv.org/abs/0907.1636}{arXiv:0907.1636 [hep-th]}\BibitemShut
  {NoStop}%
%%CITATION = ARXIV:0907.1636;%%
\bibitem [{\citenamefont {Koyama}\ and\ \citenamefont
  {Arroja}(2010)}]{Koyama:2009hc}%
  \BibitemOpen
  \bibfield  {author} {\bibinfo {author} {\bibfnamefont {K.}~\bibnamefont
  {Koyama}}\ and\ \bibinfo {author} {\bibfnamefont {F.}~\bibnamefont
  {Arroja}},\ }\bibfield  {title} {\enquote {\bibinfo {title} {{Pathological
  behaviour of the scalar graviton in Ho\v{r}ava-Lifshitz gravity}},}\ }\href
  {\doibase 10.1007/JHEP03(2010)061} {\bibfield  {journal} {\bibinfo  {journal}
  {JHEP}\ }\textbf {\bibinfo {volume} {03}},\ \bibinfo {pages} {061} (\bibinfo
  {year} {2010})},\ \Eprint {http://arxiv.org/abs/0910.1998}{arXiv:0910.1998
  [hep-th]}\BibitemShut {NoStop}%
\bibitem [{\citenamefont {Wang}(2017)}]{Wang:2017brl}%
  \BibitemOpen
  \bibfield  {author} {\bibinfo {author} {\bibfnamefont {A.}~\bibnamefont
  {Wang}},\ }\bibfield  {title} {\enquote {\bibinfo {title} {{Ho\v{r}ava
  gravity at a Lifshitz point: A progress report}},}\ }\href {\doibase
  10.1142/S0218271817300142} {\bibfield  {journal} {\bibinfo  {journal} {Int.
  J. Mod. Phys. D}\ }\textbf {\bibinfo {volume} {26}},\ \bibinfo {pages}
  {1730014} (\bibinfo {year} {2017})},\ \Eprint
  {http://arxiv.org/abs/1701.06087}{arXiv:1701.06087 [gr-qc]}\BibitemShut
  {NoStop}%
\bibitem [{\citenamefont {Audren}\ \emph {et~al.}(2015)\citenamefont {Audren},
  \citenamefont {Blas}, \citenamefont {Ivanov}, \citenamefont {Lesgourgues},\
  and\ \citenamefont {Sibiryakov}}]{Audren:2014hza}%
  \BibitemOpen
  \bibfield  {author} {\bibinfo {author} {\bibfnamefont {B.}~\bibnamefont
  {Audren}}, \bibinfo {author} {\bibfnamefont {D.}~\bibnamefont {Blas}},
  \bibinfo {author} {\bibfnamefont {M.~M.}\ \bibnamefont {Ivanov}}, \bibinfo
  {author} {\bibfnamefont {J.}~\bibnamefont {Lesgourgues}}, \ and\ \bibinfo
  {author} {\bibfnamefont {S.}~\bibnamefont {Sibiryakov}},\ }\bibfield  {title}
  {\enquote {\bibinfo {title} {{Cosmological constraints on deviations from
  Lorentz invariance in gravity and dark matter}},}\ }\href {\doibase
  10.1088/1475-7516/2015/03/016} {\bibfield  {journal} {\bibinfo  {journal}
  {JCAP}\ }\textbf {\bibinfo {volume} {1503}},\ \bibinfo {pages} {016}
  (\bibinfo {year} {2015})},\ \Eprint
  {http://arxiv.org/abs/1410.6514}{arXiv:1410.6514 [astro-ph.CO]}\BibitemShut
  {NoStop}%
%%CITATION = ARXIV:1410.6514;%%
\bibitem [{\citenamefont {Blas}\ \emph
  {et~al.}(2010{\natexlab{a}})\citenamefont {Blas}, \citenamefont {Pujolas},\
  and\ \citenamefont {Sibiryakov}}]{Blas:2009qj}%
  \BibitemOpen
  \bibfield  {author} {\bibinfo {author} {\bibfnamefont {D.}~\bibnamefont
  {Blas}}, \bibinfo {author} {\bibfnamefont {O.}~\bibnamefont {Pujolas}}, \
  and\ \bibinfo {author} {\bibfnamefont {S.}~\bibnamefont {Sibiryakov}},\
  }\bibfield  {title} {\enquote {\bibinfo {title} {{Consistent Extension of
  Hořava Gravity}},}\ }\href {\doibase 10.1103/PhysRevLett.104.181302}
  {\bibfield  {journal} {\bibinfo  {journal} {Phys. Rev. Lett.}\ }\textbf
  {\bibinfo {volume} {104}},\ \bibinfo {pages} {181302} (\bibinfo {year}
  {2010}{\natexlab{a}})},\ \Eprint
  {http://arxiv.org/abs/0909.3525}{arXiv:0909.3525 [hep-th]}\BibitemShut
  {NoStop}%
%%CITATION = ARXIV:0909.3525;%%
\bibitem [{\citenamefont {Blas}\ and\ \citenamefont
  {Sanctuary}(2011)}]{Blas:2011zd}%
  \BibitemOpen
  \bibfield  {author} {\bibinfo {author} {\bibfnamefont {D.}~\bibnamefont
  {Blas}}\ and\ \bibinfo {author} {\bibfnamefont {H.}~\bibnamefont
  {Sanctuary}},\ }\bibfield  {title} {\enquote {\bibinfo {title}
  {{Gravitational Radiation in Hořava Gravity}},}\ }\href {\doibase
  10.1103/PhysRevD.84.064004} {\bibfield  {journal} {\bibinfo  {journal} {Phys.
  Rev.}\ }\textbf {\bibinfo {volume} {D84}},\ \bibinfo {pages} {064004}
  (\bibinfo {year} {2011})},\ \Eprint
  {http://arxiv.org/abs/1105.5149}{arXiv:1105.5149 [gr-qc]}\BibitemShut
  {NoStop}%
%%CITATION = ARXIV:1105.5149;%%
\bibitem [{\citenamefont {Calcagni}(2009)}]{Calcagni:2009ar}%
  \BibitemOpen
  \bibfield  {author} {\bibinfo {author} {\bibfnamefont {G.}~\bibnamefont
  {Calcagni}},\ }\bibfield  {title} {\enquote {\bibinfo {title} {{Cosmology of
  the Lifshitz universe}},}\ }\href {\doibase 10.1088/1126-6708/2009/09/112}
  {\bibfield  {journal} {\bibinfo  {journal} {JHEP}\ }\textbf {\bibinfo
  {volume} {09}},\ \bibinfo {pages} {112} (\bibinfo {year} {2009})},\ \Eprint
  {http://arxiv.org/abs/0904.0829}{arXiv:0904.0829 [hep-th]}\BibitemShut
  {NoStop}%
%%CITATION = ARXIV:0904.0829;%%
\bibitem [{\citenamefont {Colombo}\ \emph
  {et~al.}(2015{\natexlab{a}})\citenamefont {Colombo}, \citenamefont
  {Gümrükçüoğlu},\ and\ \citenamefont {Sotiriou}}]{Colombo:2015yha}%
  \BibitemOpen
  \bibfield  {author} {\bibinfo {author} {\bibfnamefont {M.}~\bibnamefont
  {Colombo}}, \bibinfo {author} {\bibfnamefont {A.~E.}\ \bibnamefont
  {Gümrükçüoğlu}}, \ and\ \bibinfo {author} {\bibfnamefont {T.~P.}\
  \bibnamefont {Sotiriou}},\ }\bibfield  {title} {\enquote {\bibinfo {title}
  {{Hořava gravity with mixed derivative terms: Power counting
  renormalizability with lower order dispersions}},}\ }\href {\doibase
  10.1103/PhysRevD.92.064037} {\bibfield  {journal} {\bibinfo  {journal} {Phys.
  Rev.}\ }\textbf {\bibinfo {volume} {D92}},\ \bibinfo {pages} {064037}
  (\bibinfo {year} {2015}{\natexlab{a}})},\ \Eprint
  {http://arxiv.org/abs/1503.07544}{arXiv:1503.07544 [hep-th]}\BibitemShut
  {NoStop}%
%%CITATION = ARXIV:1503.07544;%%
\bibitem [{\citenamefont {Czuchry}(2011{\natexlab{a}})}]{Czuchry:2009hz}%
  \BibitemOpen
  \bibfield  {author} {\bibinfo {author} {\bibfnamefont {E.}~\bibnamefont
  {Czuchry}},\ }\bibfield  {title} {\enquote {\bibinfo {title} {{The Phase
  portrait of a matter bounce in Hořava-Lifshitz cosmology}},}\ }\href
  {\doibase 10.1088/0264-9381/28/8/085011} {\bibfield  {journal} {\bibinfo
  {journal} {Class. Quant. Grav.}\ }\textbf {\bibinfo {volume} {28}},\ \bibinfo
  {pages} {085011} (\bibinfo {year} {2011}{\natexlab{a}})},\ \Eprint
  {http://arxiv.org/abs/0911.3891}{arXiv:0911.3891 [hep-th]}\BibitemShut
  {NoStop}%
%%CITATION = ARXIV:0911.3891;%%
\bibitem [{\citenamefont {Czuchry}(2011{\natexlab{b}})}]{Czuchry:2010vx}%
  \BibitemOpen
  \bibfield  {author} {\bibinfo {author} {\bibfnamefont {E.}~\bibnamefont
  {Czuchry}},\ }\bibfield  {title} {\enquote {\bibinfo {title} {{Bounce
  scenarios in the Sotiriou-Visser-Weinfurtner generalization of the
  projectable Hořava-Lifshitz gravity}},}\ }\href {\doibase
  10.1088/0264-9381/28/12/125013} {\bibfield  {journal} {\bibinfo  {journal}
  {Class. Quant. Grav.}\ }\textbf {\bibinfo {volume} {28}},\ \bibinfo {pages}
  {125013} (\bibinfo {year} {2011}{\natexlab{b}})},\ \Eprint
  {http://arxiv.org/abs/1008.3410}{arXiv:1008.3410 [hep-th]}\BibitemShut
  {NoStop}%
%%CITATION = ARXIV:1008.3410;%%
\bibitem [{\citenamefont {Dutta}\ and\ \citenamefont
  {Saridakis}(2010{\natexlab{a}})}]{Dutta:2009jn}%
  \BibitemOpen
  \bibfield  {author} {\bibinfo {author} {\bibfnamefont {S.}~\bibnamefont
  {Dutta}}\ and\ \bibinfo {author} {\bibfnamefont {E.~N.}\ \bibnamefont
  {Saridakis}},\ }\bibfield  {title} {\enquote {\bibinfo {title}
  {{Observational constraints on Hořava-Lifshitz cosmology}},}\ }\href
  {\doibase 10.1088/1475-7516/2010/01/013} {\bibfield  {journal} {\bibinfo
  {journal} {JCAP}\ }\textbf {\bibinfo {volume} {1001}},\ \bibinfo {pages}
  {013} (\bibinfo {year} {2010}{\natexlab{a}})},\ \Eprint
  {http://arxiv.org/abs/0911.1435}{arXiv:0911.1435 [hep-th]}\BibitemShut
  {NoStop}%
%%CITATION = ARXIV:0911.1435;%%
\bibitem [{\citenamefont {Dutta}\ and\ \citenamefont
  {Saridakis}(2010{\natexlab{b}})}]{Dutta:2010jh}%
  \BibitemOpen
  \bibfield  {author} {\bibinfo {author} {\bibfnamefont {S.}~\bibnamefont
  {Dutta}}\ and\ \bibinfo {author} {\bibfnamefont {E.~N.}\ \bibnamefont
  {Saridakis}},\ }\bibfield  {title} {\enquote {\bibinfo {title} {{Overall
  observational constraints on the running parameter $\lambda$ of
  Horava-Lifshitz gravity}},}\ }\href {\doibase 10.1088/1475-7516/2010/05/013}
  {\bibfield  {journal} {\bibinfo  {journal} {JCAP}\ }\textbf {\bibinfo
  {volume} {1005}},\ \bibinfo {pages} {013} (\bibinfo {year}
  {2010}{\natexlab{b}})},\ \Eprint
  {http://arxiv.org/abs/1002.3373}{arXiv:1002.3373 [hep-th]}\BibitemShut
  {NoStop}%
%%CITATION = ARXIV:1002.3373;%%
\bibitem [{\citenamefont {Frusciante}\ \emph {et~al.}(2016)\citenamefont
  {Frusciante}, \citenamefont {Raveri}, \citenamefont {Vernieri}, \citenamefont
  {Hu},\ and\ \citenamefont {Silvestri}}]{Frusciante:2015maa}%
  \BibitemOpen
  \bibfield  {author} {\bibinfo {author} {\bibfnamefont {N.}~\bibnamefont
  {Frusciante}}, \bibinfo {author} {\bibfnamefont {M.}~\bibnamefont {Raveri}},
  \bibinfo {author} {\bibfnamefont {D.}~\bibnamefont {Vernieri}}, \bibinfo
  {author} {\bibfnamefont {B.}~\bibnamefont {Hu}}, \ and\ \bibinfo {author}
  {\bibfnamefont {A.}~\bibnamefont {Silvestri}},\ }\bibfield  {title} {\enquote
  {\bibinfo {title} {{Hořava Gravity in the Effective Field Theory formalism:
  From cosmology to observational constraints}},}\ }\href {\doibase
  10.1016/j.dark.2016.03.002} {\bibfield  {journal} {\bibinfo  {journal} {Phys.
  Dark Univ.}\ }\textbf {\bibinfo {volume} {13}},\ \bibinfo {pages} {7}
  (\bibinfo {year} {2016})},\ \Eprint
  {http://arxiv.org/abs/1508.01787}{arXiv:1508.01787 [astro-ph.CO]}\BibitemShut
  {NoStop}%
%%CITATION = ARXIV:1508.01787;%%
\bibitem [{\citenamefont {Kiritsis}\ and\ \citenamefont
  {Kofinas}(2009)}]{Kiritsis:2009sh}%
  \BibitemOpen
  \bibfield  {author} {\bibinfo {author} {\bibfnamefont {E.}~\bibnamefont
  {Kiritsis}}\ and\ \bibinfo {author} {\bibfnamefont {G.}~\bibnamefont
  {Kofinas}},\ }\bibfield  {title} {\enquote {\bibinfo {title}
  {{Hořava-Lifshitz Cosmology}},}\ }\href {\doibase
  10.1016/j.nuclphysb.2009.05.005} {\bibfield  {journal} {\bibinfo  {journal}
  {Nucl. Phys.}\ }\textbf {\bibinfo {volume} {B821}},\ \bibinfo {pages} {467}
  (\bibinfo {year} {2009})},\ \Eprint
  {http://arxiv.org/abs/0904.1334}{arXiv:0904.1334 [hep-th]}\BibitemShut
  {NoStop}%
%%CITATION = ARXIV:0904.1334;%%
\bibitem [{\citenamefont {{L{\"u}}}\ \emph {et~al.}(2009)\citenamefont
  {{L{\"u}}}, \citenamefont {{Mei}},\ and\ \citenamefont {{Pope}}}]{Lu:2009em}%
  \BibitemOpen
  \bibfield  {author} {\bibinfo {author} {\bibfnamefont {H.}~\bibnamefont
  {{L{\"u}}}}, \bibinfo {author} {\bibfnamefont {J.}~\bibnamefont {{Mei}}}, \
  and\ \bibinfo {author} {\bibfnamefont {C.~N.}\ \bibnamefont {{Pope}}},\
  }\bibfield  {title} {\enquote {\bibinfo {title} {Solutions to ho{\v r}ava
  gravity},}\ }\href {\doibase 10.1103/PhysRevLett.103.091301} {\bibfield
  {journal} {\bibinfo  {journal} {Phys. Rev. Lett.}\ }\textbf {\bibinfo
  {volume} {103}},\ \bibinfo {pages} {091301} (\bibinfo {year} {2009})},\
  \Eprint {http://arxiv.org/abs/0904.1595}{arXiv:0904.1595
  [hep-th]}\BibitemShut {NoStop}%
%%CITATION = ARXIV:0904.1595;%%
\bibitem [{\citenamefont {Saridakis}(2010)}]{Saridakis:2009bv}%
  \BibitemOpen
  \bibfield  {author} {\bibinfo {author} {\bibfnamefont {E.~N.}\ \bibnamefont
  {Saridakis}},\ }\bibfield  {title} {\enquote {\bibinfo {title}
  {{Hořava-Lifshitz Dark Energy}},}\ }\href {\doibase
  10.1140/epjc/s10052-010-1294-6} {\bibfield  {journal} {\bibinfo  {journal}
  {Eur. Phys. J.}\ }\textbf {\bibinfo {volume} {C67}},\ \bibinfo {pages} {229}
  (\bibinfo {year} {2010})},\ \Eprint
  {http://arxiv.org/abs/0905.3532}{arXiv:0905.3532 [hep-th]}\BibitemShut
  {NoStop}%
%%CITATION = ARXIV:0905.3532;%%
\bibitem [{\citenamefont {Park}(2010)}]{Park:2009zr}%
  \BibitemOpen
  \bibfield  {author} {\bibinfo {author} {\bibfnamefont {M.-i.}\ \bibnamefont
  {Park}},\ }\bibfield  {title} {\enquote {\bibinfo {title} {{A Test of Horava
  Gravity: The Dark Energy}},}\ }\href {\doibase 10.1088/1475-7516/2010/01/001}
  {\bibfield  {journal} {\bibinfo  {journal} {JCAP}\ }\textbf {\bibinfo
  {volume} {01}},\ \bibinfo {pages} {001} (\bibinfo {year} {2010})},\ \Eprint
  {http://arxiv.org/abs/0906.4275}{arXiv:0906.4275 [hep-th]}\BibitemShut
  {NoStop}%
\bibitem [{\citenamefont {Park}(2009)}]{Park:2009zra}%
  \BibitemOpen
  \bibfield  {author} {\bibinfo {author} {\bibfnamefont {M.-i.}\ \bibnamefont
  {Park}},\ }\bibfield  {title} {\enquote {\bibinfo {title} {{The Black Hole
  and Cosmological Solutions in IR modified Horava Gravity}},}\ }\href
  {\doibase 10.1088/1126-6708/2009/09/123} {\bibfield  {journal} {\bibinfo
  {journal} {JHEP}\ }\textbf {\bibinfo {volume} {09}},\ \bibinfo {pages} {123}
  (\bibinfo {year} {2009})},\ \Eprint
  {http://arxiv.org/abs/0905.4480}{arXiv:0905.4480 [hep-th]}\BibitemShut
  {NoStop}%
\bibitem [{\citenamefont {Li}\ and\ \citenamefont {He}(2021)}]{Li:2021riw}%
  \BibitemOpen
  \bibfield  {author} {\bibinfo {author} {\bibfnamefont {G.-P.}\ \bibnamefont
  {Li}}\ and\ \bibinfo {author} {\bibfnamefont {K.-J.}\ \bibnamefont {He}},\
  }\bibfield  {title} {\enquote {\bibinfo {title} {{Shadows and rings of the
  Kehagias-Sfetsos black hole surrounded by thin disk accretion}},}\ }\href
  {\doibase 10.1088/1475-7516/2021/06/037} {\bibfield  {journal} {\bibinfo
  {journal} {JCAP}\ }\textbf {\bibinfo {volume} {06}},\ \bibinfo {pages} {037}
  (\bibinfo {year} {2021})},\ \Eprint
  {http://arxiv.org/abs/2105.08521}{arXiv:2105.08521 [gr-qc]}\BibitemShut
  {NoStop}%
\bibitem [{\citenamefont {Jusufi}\ \emph {et~al.}(2022)\citenamefont {Jusufi},
  \citenamefont {Hassanabadi}, \citenamefont {Sedaghatnia}, \citenamefont
  {Kr\'\i{}z}, \citenamefont {Chung}, \citenamefont {Chen}, \citenamefont
  {Zhao},\ and\ \citenamefont {Long}}]{Jusufi:2022ava}%
  \BibitemOpen
  \bibfield  {author} {\bibinfo {author} {\bibfnamefont {K.}~\bibnamefont
  {Jusufi}}, \bibinfo {author} {\bibfnamefont {H.}~\bibnamefont {Hassanabadi}},
  \bibinfo {author} {\bibfnamefont {P.}~\bibnamefont {Sedaghatnia}}, \bibinfo
  {author} {\bibfnamefont {J.}~\bibnamefont {Kr\'\i{}z}}, \bibinfo {author}
  {\bibfnamefont {W.~S.}\ \bibnamefont {Chung}}, \bibinfo {author}
  {\bibfnamefont {H.}~\bibnamefont {Chen}}, \bibinfo {author} {\bibfnamefont
  {Z.-L.}\ \bibnamefont {Zhao}}, \ and\ \bibinfo {author} {\bibfnamefont
  {Z.~W.}\ \bibnamefont {Long}},\ }\bibfield  {title} {\enquote {\bibinfo
  {title} {{Thermodynamics and shadow images of charged black holes in
  Horava\textendash{}Lifshitz gravity}},}\ }\href {\doibase
  10.1140/epjp/s13360-022-03354-7} {\bibfield  {journal} {\bibinfo  {journal}
  {Eur. Phys. J. Plus}\ }\textbf {\bibinfo {volume} {137}},\ \bibinfo {pages}
  {1147} (\bibinfo {year} {2022})}\BibitemShut {NoStop}%
\bibitem [{\citenamefont {Gong}\ \emph {et~al.}(2022)\citenamefont {Gong},
  \citenamefont {Zhu}, \citenamefont {Niu}, \citenamefont {Wu}, \citenamefont
  {Cui}, \citenamefont {Zhang}, \citenamefont {Zhao},\ and\ \citenamefont
  {Wang}}]{Gong:2021jgg}%
  \BibitemOpen
  \bibfield  {author} {\bibinfo {author} {\bibfnamefont {C.}~\bibnamefont
  {Gong}}, \bibinfo {author} {\bibfnamefont {T.}~\bibnamefont {Zhu}}, \bibinfo
  {author} {\bibfnamefont {R.}~\bibnamefont {Niu}}, \bibinfo {author}
  {\bibfnamefont {Q.}~\bibnamefont {Wu}}, \bibinfo {author} {\bibfnamefont
  {J.-L.}\ \bibnamefont {Cui}}, \bibinfo {author} {\bibfnamefont
  {X.}~\bibnamefont {Zhang}}, \bibinfo {author} {\bibfnamefont
  {W.}~\bibnamefont {Zhao}}, \ and\ \bibinfo {author} {\bibfnamefont
  {A.}~\bibnamefont {Wang}},\ }\bibfield  {title} {\enquote {\bibinfo {title}
  {{Gravitational wave constraints on Lorentz and parity violations in gravity:
  High-order spatial derivative cases}},}\ }\href {\doibase
  10.1103/PhysRevD.105.044034} {\bibfield  {journal} {\bibinfo  {journal}
  {Phys. Rev. D}\ }\textbf {\bibinfo {volume} {105}},\ \bibinfo {pages}
  {044034} (\bibinfo {year} {2022})},\ \Eprint
  {http://arxiv.org/abs/2112.06446}{arXiv:2112.06446 [gr-qc]}\BibitemShut
  {NoStop}%
\bibitem [{\citenamefont {Emir~G\"umr\"uk\c{c}\"uo\u{g}lu}\ \emph
  {et~al.}(2018)\citenamefont {Emir~G\"umr\"uk\c{c}\"uo\u{g}lu}, \citenamefont
  {Saravani},\ and\ \citenamefont {Sotiriou}}]{EmirGumrukcuoglu:2017cfa}%
  \BibitemOpen
  \bibfield  {author} {\bibinfo {author} {\bibfnamefont {A.}~\bibnamefont
  {Emir~G\"umr\"uk\c{c}\"uo\u{g}lu}}, \bibinfo {author} {\bibfnamefont
  {M.}~\bibnamefont {Saravani}}, \ and\ \bibinfo {author} {\bibfnamefont
  {T.~P.}\ \bibnamefont {Sotiriou}},\ }\bibfield  {title} {\enquote {\bibinfo
  {title} {{Ho\v{r}ava gravity after GW170817}},}\ }\href {\doibase
  10.1103/PhysRevD.97.024032} {\bibfield  {journal} {\bibinfo  {journal} {Phys.
  Rev. D}\ }\textbf {\bibinfo {volume} {97}},\ \bibinfo {pages} {024032}
  (\bibinfo {year} {2018})},\ \Eprint
  {http://arxiv.org/abs/1711.08845}{arXiv:1711.08845 [gr-qc]}\BibitemShut
  {NoStop}%
\bibitem [{\citenamefont {Sotiriou}(2011)}]{Sotiriou:2010wn}%
  \BibitemOpen
  \bibfield  {author} {\bibinfo {author} {\bibfnamefont {T.~P.}\ \bibnamefont
  {Sotiriou}},\ }\bibfield  {title} {\enquote {\bibinfo {title}
  {{Hořava-Lifshitz gravity: a status report}},}\ }\bibfield  {booktitle}
  {\emph {\bibinfo {booktitle} {{Proceedings, 14th Conference on Recent
  developments in gravity (NEB 14): Ioannina, Greece, June 8-11, 2010}}},\
  }\href {\doibase 10.1088/1742-6596/283/1/012034} {\bibfield  {journal}
  {\bibinfo  {journal} {J. Phys. Conf. Ser.}\ }\textbf {\bibinfo {volume}
  {283}},\ \bibinfo {pages} {012034} (\bibinfo {year} {2011})},\ \Eprint
  {http://arxiv.org/abs/1010.3218}{arXiv:1010.3218 [hep-th]}\BibitemShut
  {NoStop}%
%%CITATION = ARXIV:1010.3218;%%
\bibitem [{\citenamefont {Blas}\ \emph {et~al.}(2009)\citenamefont {Blas},
  \citenamefont {Pujolas},\ and\ \citenamefont {Sibiryakov}}]{Blas:2009yd}%
  \BibitemOpen
  \bibfield  {author} {\bibinfo {author} {\bibfnamefont {D.}~\bibnamefont
  {Blas}}, \bibinfo {author} {\bibfnamefont {O.}~\bibnamefont {Pujolas}}, \
  and\ \bibinfo {author} {\bibfnamefont {S.}~\bibnamefont {Sibiryakov}},\
  }\bibfield  {title} {\enquote {\bibinfo {title} {{On the Extra Mode and
  Inconsistency of Horava Gravity}},}\ }\href {\doibase
  10.1088/1126-6708/2009/10/029} {\bibfield  {journal} {\bibinfo  {journal}
  {JHEP}\ }\textbf {\bibinfo {volume} {10}},\ \bibinfo {pages} {029} (\bibinfo
  {year} {2009})},\ \Eprint {http://arxiv.org/abs/0906.3046}{arXiv:0906.3046
  [hep-th]}\BibitemShut {NoStop}%
\bibitem [{\citenamefont {Charmousis}\ \emph {et~al.}(2009)\citenamefont
  {Charmousis}, \citenamefont {Niz}, \citenamefont {Padilla},\ and\
  \citenamefont {Saffin}}]{Charmousis:2009tc}%
  \BibitemOpen
  \bibfield  {author} {\bibinfo {author} {\bibfnamefont {C.}~\bibnamefont
  {Charmousis}}, \bibinfo {author} {\bibfnamefont {G.}~\bibnamefont {Niz}},
  \bibinfo {author} {\bibfnamefont {A.}~\bibnamefont {Padilla}}, \ and\
  \bibinfo {author} {\bibfnamefont {P.~M.}\ \bibnamefont {Saffin}},\ }\bibfield
   {title} {\enquote {\bibinfo {title} {{Strong coupling in Hořava
  gravity}},}\ }\href {\doibase 10.1088/1126-6708/2009/08/070} {\bibfield
  {journal} {\bibinfo  {journal} {JHEP}\ }\textbf {\bibinfo {volume} {08}},\
  \bibinfo {pages} {070} (\bibinfo {year} {2009})},\ \Eprint
  {http://arxiv.org/abs/0905.2579}{arXiv:0905.2579 [hep-th]}\BibitemShut
  {NoStop}%
%%CITATION = ARXIV:0905.2579;%%
\bibitem [{\citenamefont {Vernieri}\ and\ \citenamefont
  {Sotiriou}(2012)}]{Vernieri:2011aa}%
  \BibitemOpen
  \bibfield  {author} {\bibinfo {author} {\bibfnamefont {D.}~\bibnamefont
  {Vernieri}}\ and\ \bibinfo {author} {\bibfnamefont {T.~P.}\ \bibnamefont
  {Sotiriou}},\ }\bibfield  {title} {\enquote {\bibinfo {title}
  {{Hořava-Lifshitz Gravity: Detailed Balance Revisited}},}\ }\href {\doibase
  10.1103/PhysRevD.85.069901, 10.1103/PhysRevD.85.064003} {\bibfield  {journal}
  {\bibinfo  {journal} {Phys. Rev.}\ }\textbf {\bibinfo {volume} {D85}},\
  \bibinfo {pages} {064003} (\bibinfo {year} {2012})},\ \Eprint
  {http://arxiv.org/abs/1112.3385}{arXiv:1112.3385 [hep-th]}\BibitemShut
  {NoStop}%
%%CITATION = ARXIV:1112.3385;%%
\bibitem [{\citenamefont {Appignani}\ \emph {et~al.}(2010)\citenamefont
  {Appignani}, \citenamefont {Casadio},\ and\ \citenamefont
  {Shankaranarayanan}}]{Appignani:2009dy}%
  \BibitemOpen
  \bibfield  {author} {\bibinfo {author} {\bibfnamefont {C.}~\bibnamefont
  {Appignani}}, \bibinfo {author} {\bibfnamefont {R.}~\bibnamefont {Casadio}},
  \ and\ \bibinfo {author} {\bibfnamefont {S.}~\bibnamefont
  {Shankaranarayanan}},\ }\bibfield  {title} {\enquote {\bibinfo {title} {{The
  Cosmological Constant and Horava-Lifshitz Gravity}},}\ }\href {\doibase
  10.1088/1475-7516/2010/04/006} {\bibfield  {journal} {\bibinfo  {journal}
  {JCAP}\ }\textbf {\bibinfo {volume} {04}},\ \bibinfo {pages} {006} (\bibinfo
  {year} {2010})},\ \Eprint {http://arxiv.org/abs/0907.3121}{arXiv:0907.3121
  [hep-th]}\BibitemShut {NoStop}%
\bibitem [{\citenamefont {Vernieri}(2015)}]{Vernieri:2015uma}%
  \BibitemOpen
  \bibfield  {author} {\bibinfo {author} {\bibfnamefont {D.}~\bibnamefont
  {Vernieri}},\ }\bibfield  {title} {\enquote {\bibinfo {title} {{On
  power-counting renormalizability of Ho\v{r}ava gravity with detailed
  balance}},}\ }\href {\doibase 10.1103/PhysRevD.91.124029} {\bibfield
  {journal} {\bibinfo  {journal} {Phys. Rev. D}\ }\textbf {\bibinfo {volume}
  {91}},\ \bibinfo {pages} {124029} (\bibinfo {year} {2015})},\ \Eprint
  {http://arxiv.org/abs/1502.06607}{arXiv:1502.06607 [hep-th]}\BibitemShut
  {NoStop}%
\bibitem [{\citenamefont {Bertolami}\ and\ \citenamefont
  {Zarro}(2011)}]{Bertolami:2011ka}%
  \BibitemOpen
  \bibfield  {author} {\bibinfo {author} {\bibfnamefont {O.}~\bibnamefont
  {Bertolami}}\ and\ \bibinfo {author} {\bibfnamefont {C.~A.~D.}\ \bibnamefont
  {Zarro}},\ }\bibfield  {title} {\enquote {\bibinfo {title}
  {{Ho\v{r}ava-Lifshitz Quantum Cosmology}},}\ }\href {\doibase
  10.1103/PhysRevD.84.044042} {\bibfield  {journal} {\bibinfo  {journal} {Phys.
  Rev. D}\ }\textbf {\bibinfo {volume} {84}},\ \bibinfo {pages} {044042}
  (\bibinfo {year} {2011})},\ \Eprint
  {http://arxiv.org/abs/1106.0126}{arXiv:1106.0126 [hep-th]}\BibitemShut
  {NoStop}%
\bibitem [{\citenamefont {Christodoulakis}\ and\ \citenamefont
  {Dimakis}(2012)}]{Christodoulakis:2011np}%
  \BibitemOpen
  \bibfield  {author} {\bibinfo {author} {\bibfnamefont {T.}~\bibnamefont
  {Christodoulakis}}\ and\ \bibinfo {author} {\bibfnamefont {N.}~\bibnamefont
  {Dimakis}},\ }\bibfield  {title} {\enquote {\bibinfo {title} {{Classical and
  Quantum Bianchi Type III vacuum Horava-Lifshitz Cosmology}},}\ }\href
  {\doibase 10.1016/j.geomphys.2012.09.005} {\bibfield  {journal} {\bibinfo
  {journal} {J. Geom. Phys.}\ }\textbf {\bibinfo {volume} {62}},\ \bibinfo
  {pages} {2401} (\bibinfo {year} {2012})},\ \Eprint
  {http://arxiv.org/abs/1112.0903}{arXiv:1112.0903 [gr-qc]}\BibitemShut
  {NoStop}%
\bibitem [{\citenamefont {Pitelli}\ and\ \citenamefont
  {Saa}(2012)}]{Pitelli:2012sj}%
  \BibitemOpen
  \bibfield  {author} {\bibinfo {author} {\bibfnamefont {J.~P.~M.}\
  \bibnamefont {Pitelli}}\ and\ \bibinfo {author} {\bibfnamefont
  {A.}~\bibnamefont {Saa}},\ }\bibfield  {title} {\enquote {\bibinfo {title}
  {{Quantum Singularities in Horava-Lifshitz Cosmology}},}\ }\href {\doibase
  10.1103/PhysRevD.86.063506} {\bibfield  {journal} {\bibinfo  {journal} {Phys.
  Rev. D}\ }\textbf {\bibinfo {volume} {86}},\ \bibinfo {pages} {063506}
  (\bibinfo {year} {2012})},\ \Eprint
  {http://arxiv.org/abs/1204.4924}{arXiv:1204.4924 [gr-qc]}\BibitemShut
  {NoStop}%
\bibitem [{\citenamefont {Vakili}\ and\ \citenamefont
  {Kord}(2013)}]{Vakili:2013wc}%
  \BibitemOpen
  \bibfield  {author} {\bibinfo {author} {\bibfnamefont {B.}~\bibnamefont
  {Vakili}}\ and\ \bibinfo {author} {\bibfnamefont {V.}~\bibnamefont {Kord}},\
  }\bibfield  {title} {\enquote {\bibinfo {title} {{Classical and quantum
  Ho\v{r}ava-Lifshitz cosmology in a minisuperspace perspective}},}\ }\href
  {\doibase 10.1007/s10714-013-1527-8} {\bibfield  {journal} {\bibinfo
  {journal} {Gen. Rel. Grav.}\ }\textbf {\bibinfo {volume} {45}},\ \bibinfo
  {pages} {1313} (\bibinfo {year} {2013})},\ \Eprint
  {http://arxiv.org/abs/1301.0809}{arXiv:1301.0809 [gr-qc]}\BibitemShut
  {NoStop}%
\bibitem [{\citenamefont {Obregon}\ and\ \citenamefont
  {Preciado}(2012)}]{Obregon:2012bt}%
  \BibitemOpen
  \bibfield  {author} {\bibinfo {author} {\bibfnamefont {O.}~\bibnamefont
  {Obregon}}\ and\ \bibinfo {author} {\bibfnamefont {J.~A.}\ \bibnamefont
  {Preciado}},\ }\bibfield  {title} {\enquote {\bibinfo {title} {{Quantum
  cosmology in Ho\v{r}ava-Lifshitz gravity}},}\ }\href {\doibase
  10.1103/PhysRevD.86.063502} {\bibfield  {journal} {\bibinfo  {journal} {Phys.
  Rev. D}\ }\textbf {\bibinfo {volume} {86}},\ \bibinfo {pages} {063502}
  (\bibinfo {year} {2012})},\ \Eprint
  {http://arxiv.org/abs/1305.6950}{arXiv:1305.6950 [gr-qc]}\BibitemShut
  {NoStop}%
\bibitem [{\citenamefont {Benedetti}\ and\ \citenamefont
  {Henson}(2015)}]{Benedetti:2014dra}%
  \BibitemOpen
  \bibfield  {author} {\bibinfo {author} {\bibfnamefont {D.}~\bibnamefont
  {Benedetti}}\ and\ \bibinfo {author} {\bibfnamefont {J.}~\bibnamefont
  {Henson}},\ }\bibfield  {title} {\enquote {\bibinfo {title} {{Spacetime
  condensation in (2+1)-dimensional CDT from a Ho\v{r}ava\textendash{}Lifshitz
  minisuperspace model}},}\ }\href {\doibase 10.1088/0264-9381/32/21/215007}
  {\bibfield  {journal} {\bibinfo  {journal} {Class. Quant. Grav.}\ }\textbf
  {\bibinfo {volume} {32}},\ \bibinfo {pages} {215007} (\bibinfo {year}
  {2015})},\ \Eprint {http://arxiv.org/abs/1410.0845}{arXiv:1410.0845
  [gr-qc]}\BibitemShut {NoStop}%
\bibitem [{\citenamefont {Garc\'\i{}a-Compe\'an}\ and\ \citenamefont
  {Mata-Pacheco}(2022)}]{Garcia-Compean:2021vcy}%
  \BibitemOpen
  \bibfield  {author} {\bibinfo {author} {\bibfnamefont {H.}~\bibnamefont
  {Garc\'\i{}a-Compe\'an}}\ and\ \bibinfo {author} {\bibfnamefont
  {D.}~\bibnamefont {Mata-Pacheco}},\ }\bibfield  {title} {\enquote {\bibinfo
  {title} {{Lorentzian Vacuum Transitions in Ho\v{r}ava\textendash{}Lifshitz
  Gravity}},}\ }\href {\doibase 10.3390/universe8040237} {\bibfield  {journal}
  {\bibinfo  {journal} {Universe}\ }\textbf {\bibinfo {volume} {8}},\ \bibinfo
  {pages} {237} (\bibinfo {year} {2022})},\ \Eprint
  {http://arxiv.org/abs/2111.11571}{arXiv:2111.11571 [gr-qc]}\BibitemShut
  {NoStop}%
\bibitem [{\citenamefont {Papazoglou}\ and\ \citenamefont
  {Sotiriou}(2010)}]{Papazoglou:2009fj}%
  \BibitemOpen
  \bibfield  {author} {\bibinfo {author} {\bibfnamefont {A.}~\bibnamefont
  {Papazoglou}}\ and\ \bibinfo {author} {\bibfnamefont {T.~P.}\ \bibnamefont
  {Sotiriou}},\ }\bibfield  {title} {\enquote {\bibinfo {title} {{Strong
  coupling in extended Horava-Lifshitz gravity}},}\ }\href {\doibase
  10.1016/j.physletb.2010.01.054} {\bibfield  {journal} {\bibinfo  {journal}
  {Phys. Lett. B}\ }\textbf {\bibinfo {volume} {685}},\ \bibinfo {pages} {197}
  (\bibinfo {year} {2010})},\ \Eprint
  {http://arxiv.org/abs/0911.1299}{arXiv:0911.1299 [hep-th]}\BibitemShut
  {NoStop}%
\bibitem [{\citenamefont {Blas}\ \emph
  {et~al.}(2010{\natexlab{b}})\citenamefont {Blas}, \citenamefont {Pujolas},\
  and\ \citenamefont {Sibiryakov}}]{Blas:2009ck}%
  \BibitemOpen
  \bibfield  {author} {\bibinfo {author} {\bibfnamefont {D.}~\bibnamefont
  {Blas}}, \bibinfo {author} {\bibfnamefont {O.}~\bibnamefont {Pujolas}}, \
  and\ \bibinfo {author} {\bibfnamefont {S.}~\bibnamefont {Sibiryakov}},\
  }\bibfield  {title} {\enquote {\bibinfo {title} {{Comment on `Strong coupling
  in extended Horava-Lifshitz gravity'}},}\ }\href {\doibase
  10.1016/j.physletb.2010.03.073} {\bibfield  {journal} {\bibinfo  {journal}
  {Phys. Lett. B}\ }\textbf {\bibinfo {volume} {688}},\ \bibinfo {pages} {350}
  (\bibinfo {year} {2010}{\natexlab{b}})},\ \Eprint
  {http://arxiv.org/abs/0912.0550}{arXiv:0912.0550 [hep-th]}\BibitemShut
  {NoStop}%
\bibitem [{\citenamefont {Nilsson}\ and\ \citenamefont
  {Czuchry}(2018)}]{Nilsson:2018knn}%
  \BibitemOpen
  \bibfield  {author} {\bibinfo {author} {\bibfnamefont {N.~A.}\ \bibnamefont
  {Nilsson}}\ and\ \bibinfo {author} {\bibfnamefont {E.}~\bibnamefont
  {Czuchry}},\ }\bibfield  {title} {\enquote {\bibinfo {title}
  {{Ho\v{r}ava-Lifshitz cosmology in light of new data}},}\ }\href {\doibase
  10.1016/j.dark.2018.100253} {\bibfield  {journal} {\bibinfo  {journal} {Phys.
  Dark Univ.}\ ,\ \bibinfo {pages} {100253}} (\bibinfo {year} {2018})},\
  \bibinfo {note} {[Phys. Dark Univ.C23,100253(2019)]},\ \Eprint
  {http://arxiv.org/abs/1803.03615}{arXiv:1803.03615 [gr-qc]}\BibitemShut
  {NoStop}%
%%CITATION = ARXIV:1803.03615;%%
\bibitem [{\citenamefont {Di~Valentino}\ \emph {et~al.}(2023)\citenamefont
  {Di~Valentino}, \citenamefont {Nilsson},\ and\ \citenamefont
  {Park}}]{DiValentino:2022eot}%
  \BibitemOpen
  \bibfield  {author} {\bibinfo {author} {\bibfnamefont {E.}~\bibnamefont
  {Di~Valentino}}, \bibinfo {author} {\bibfnamefont {N.~A.}\ \bibnamefont
  {Nilsson}}, \ and\ \bibinfo {author} {\bibfnamefont {M.-I.}\ \bibnamefont
  {Park}},\ }\bibfield  {title} {\enquote {\bibinfo {title} {{A new test of
  dynamical dark energy models and cosmic tensions in Ho\v{r}ava gravity}},}\
  }\href {\doibase 10.1093/mnras/stac3824} {\bibfield  {journal} {\bibinfo
  {journal} {Mon. Not. Roy. Astron. Soc.}\ }\textbf {\bibinfo {volume} {519}},\
  \bibinfo {pages} {5043} (\bibinfo {year} {2023})},\ \Eprint
  {http://arxiv.org/abs/2212.07683}{arXiv:2212.07683 [astro-ph.CO]}\BibitemShut
  {NoStop}%
\bibitem [{\citenamefont {Nilsson}\ and\ \citenamefont
  {Park}(2022)}]{Nilsson:2021ute}%
  \BibitemOpen
  \bibfield  {author} {\bibinfo {author} {\bibfnamefont {N.~A.}\ \bibnamefont
  {Nilsson}}\ and\ \bibinfo {author} {\bibfnamefont {M.-I.}\ \bibnamefont
  {Park}},\ }\bibfield  {title} {\enquote {\bibinfo {title} {{Tests of standard
  cosmology in Ho\v{r}ava gravity, Bayesian evidence for a closed universe, and
  the Hubble tension}},}\ }\href {\doibase 10.1140/epjc/s10052-022-10839-3}
  {\bibfield  {journal} {\bibinfo  {journal} {Eur. Phys. J. C}\ }\textbf
  {\bibinfo {volume} {82}},\ \bibinfo {pages} {873} (\bibinfo {year} {2022})},\
  \Eprint {http://arxiv.org/abs/2108.07986}{arXiv:2108.07986
  [hep-th]}\BibitemShut {NoStop}%
\bibitem [{\citenamefont {Nilsson}(2020)}]{Nilsson:2019bxv}%
  \BibitemOpen
  \bibfield  {author} {\bibinfo {author} {\bibfnamefont {N.~A.}\ \bibnamefont
  {Nilsson}},\ }\bibfield  {title} {\enquote {\bibinfo {title}
  {{Preferred-frame effects, the $H_0$ tension, and probes of
  Ho\v{r}ava\textendash{}Lifshitz gravity}},}\ }\href {\doibase
  10.1140/epjp/s13360-020-00369-w} {\bibfield  {journal} {\bibinfo  {journal}
  {Eur. Phys. J. Plus}\ }\textbf {\bibinfo {volume} {135}},\ \bibinfo {pages}
  {361} (\bibinfo {year} {2020})},\ \Eprint
  {http://arxiv.org/abs/1910.14414}{arXiv:1910.14414 [gr-qc]}\BibitemShut
  {NoStop}%
\bibitem [{\citenamefont {Xu}(2018)}]{Xu:2018fag}%
  \BibitemOpen
  \bibfield  {author} {\bibinfo {author} {\bibfnamefont {W.}~\bibnamefont
  {Xu}},\ }\bibfield  {title} {\enquote {\bibinfo {title} {{$\lambda$ phase
  transition in Horava gravity}},}\ }\href {\doibase 10.1155/2018/2175818}
  {\bibfield  {journal} {\bibinfo  {journal} {Adv. High Energy Phys.}\ }\textbf
  {\bibinfo {volume} {2018}},\ \bibinfo {pages} {2175818} (\bibinfo {year}
  {2018})},\ \Eprint {http://arxiv.org/abs/1804.03815}{arXiv:1804.03815
  [gr-qc]}\BibitemShut {NoStop}%
\bibitem [{\citenamefont {Handley}(2021)}]{Handley:2019tkm}%
  \BibitemOpen
  \bibfield  {author} {\bibinfo {author} {\bibfnamefont {W.}~\bibnamefont
  {Handley}},\ }\bibfield  {title} {\enquote {\bibinfo {title} {{Curvature
  tension: evidence for a closed universe}},}\ }\href {\doibase
  10.1103/PhysRevD.103.L041301} {\bibfield  {journal} {\bibinfo  {journal}
  {Phys. Rev. D}\ }\textbf {\bibinfo {volume} {103}},\ \bibinfo {pages}
  {L041301} (\bibinfo {year} {2021})},\ \Eprint
  {http://arxiv.org/abs/1908.09139}{arXiv:1908.09139 [astro-ph.CO]}\BibitemShut
  {NoStop}%
\bibitem [{\citenamefont {Hořava}(2009)}]{Horava:2009uw}%
  \BibitemOpen
  \bibfield  {author} {\bibinfo {author} {\bibfnamefont {P.}~\bibnamefont
  {Hořava}},\ }\bibfield  {title} {\enquote {\bibinfo {title} {{Quantum
  Gravity at a Lifshitz Point}},}\ }\href {\doibase 10.1103/PhysRevD.79.084008}
  {\bibfield  {journal} {\bibinfo  {journal} {Phys. Rev.}\ }\textbf {\bibinfo
  {volume} {D79}},\ \bibinfo {pages} {084008} (\bibinfo {year} {2009})},\
  \Eprint {http://arxiv.org/abs/0901.3775}{arXiv:0901.3775
  [hep-th]}\BibitemShut {NoStop}%
%%CITATION = ARXIV:0901.3775;%%
\bibitem [{\citenamefont {Kehagias}\ and\ \citenamefont
  {Sfetsos}(2009)}]{Kehagias:2009is}%
  \BibitemOpen
  \bibfield  {author} {\bibinfo {author} {\bibfnamefont {A.}~\bibnamefont
  {Kehagias}}\ and\ \bibinfo {author} {\bibfnamefont {K.}~\bibnamefont
  {Sfetsos}},\ }\bibfield  {title} {\enquote {\bibinfo {title} {{The Black hole
  and FRW geometries of non-relativistic gravity}},}\ }\href {\doibase
  10.1016/j.physletb.2009.06.019} {\bibfield  {journal} {\bibinfo  {journal}
  {Phys. Lett. B}\ }\textbf {\bibinfo {volume} {678}},\ \bibinfo {pages} {123}
  (\bibinfo {year} {2009})},\ \Eprint
  {http://arxiv.org/abs/0905.0477}{arXiv:0905.0477 [hep-th]}\BibitemShut
  {NoStop}%
\bibitem [{\citenamefont {Shin}\ and\ \citenamefont
  {Park}(2017)}]{Shin:2017ott}%
  \BibitemOpen
  \bibfield  {author} {\bibinfo {author} {\bibfnamefont {S.}~\bibnamefont
  {Shin}}\ and\ \bibinfo {author} {\bibfnamefont {M.-I.}\ \bibnamefont
  {Park}},\ }\bibfield  {title} {\enquote {\bibinfo {title} {{On gauge
  invariant cosmological perturbations in UV-modified Ho\v{r}ava gravity}},}\
  }\href {\doibase 10.1088/1475-7516/2017/12/033} {\bibfield  {journal}
  {\bibinfo  {journal} {JCAP}\ }\textbf {\bibinfo {volume} {12}},\ \bibinfo
  {pages} {033} (\bibinfo {year} {2017})},\ \Eprint
  {http://arxiv.org/abs/1701.03844}{arXiv:1701.03844 [hep-th]}\BibitemShut
  {NoStop}%
\bibitem [{\citenamefont {Kobayashi}\ \emph {et~al.}(2010)\citenamefont
  {Kobayashi}, \citenamefont {Urakawa},\ and\ \citenamefont
  {Yamaguchi}}]{Kobayashi:2010eh}%
  \BibitemOpen
  \bibfield  {author} {\bibinfo {author} {\bibfnamefont {T.}~\bibnamefont
  {Kobayashi}}, \bibinfo {author} {\bibfnamefont {Y.}~\bibnamefont {Urakawa}},
  \ and\ \bibinfo {author} {\bibfnamefont {M.}~\bibnamefont {Yamaguchi}},\
  }\bibfield  {title} {\enquote {\bibinfo {title} {{Cosmological perturbations
  in a healthy extension of Horava gravity}},}\ }\href {\doibase
  10.1088/1475-7516/2010/04/025} {\bibfield  {journal} {\bibinfo  {journal}
  {JCAP}\ }\textbf {\bibinfo {volume} {04}},\ \bibinfo {pages} {025} (\bibinfo
  {year} {2010})},\ \Eprint {http://arxiv.org/abs/1002.3101}{arXiv:1002.3101
  [hep-th]}\BibitemShut {NoStop}%
\bibitem [{\citenamefont {Carroll}\ and\ \citenamefont
  {Lim}(2004)}]{Carroll:2004ai}%
  \BibitemOpen
  \bibfield  {author} {\bibinfo {author} {\bibfnamefont {S.~M.}\ \bibnamefont
  {Carroll}}\ and\ \bibinfo {author} {\bibfnamefont {E.~A.}\ \bibnamefont
  {Lim}},\ }\bibfield  {title} {\enquote {\bibinfo {title} {{Lorentz-violating
  vector fields slow the universe down}},}\ }\href {\doibase
  10.1103/PhysRevD.70.123525} {\bibfield  {journal} {\bibinfo  {journal} {Phys.
  Rev.}\ }\textbf {\bibinfo {volume} {D70}},\ \bibinfo {pages} {123525}
  (\bibinfo {year} {2004})},\ \Eprint
  {http://arxiv.org/abs/hep-th/0407149}{arXiv:hep-th/0407149
  [hep-th]}\BibitemShut {NoStop}%
%%CITATION = HEP-TH/0407149;%%
\bibitem [{\citenamefont {Kiritsis}(2010)}]{Kiritsis:2009vz}%
  \BibitemOpen
  \bibfield  {author} {\bibinfo {author} {\bibfnamefont {E.}~\bibnamefont
  {Kiritsis}},\ }\bibfield  {title} {\enquote {\bibinfo {title} {{Spherically
  symmetric solutions in modified Horava-Lifshitz gravity}},}\ }\href {\doibase
  10.1103/PhysRevD.81.044009} {\bibfield  {journal} {\bibinfo  {journal} {Phys.
  Rev. D}\ }\textbf {\bibinfo {volume} {81}},\ \bibinfo {pages} {044009}
  (\bibinfo {year} {2010})},\ \Eprint
  {http://arxiv.org/abs/0911.3164}{arXiv:0911.3164 [hep-th]}\BibitemShut
  {NoStop}%
\bibitem [{\citenamefont {Henneaux}\ \emph {et~al.}(2010)\citenamefont
  {Henneaux}, \citenamefont {Kleinschmidt},\ and\ \citenamefont
  {Lucena~G\'omez}}]{Henneaux:2009zb}%
  \BibitemOpen
  \bibfield  {author} {\bibinfo {author} {\bibfnamefont {M.}~\bibnamefont
  {Henneaux}}, \bibinfo {author} {\bibfnamefont {A.}~\bibnamefont
  {Kleinschmidt}}, \ and\ \bibinfo {author} {\bibfnamefont {G.}~\bibnamefont
  {Lucena~G\'omez}},\ }\bibfield  {title} {\enquote {\bibinfo {title} {{A
  dynamical inconsistency of Horava gravity}},}\ }\href {\doibase
  10.1103/PhysRevD.81.064002} {\bibfield  {journal} {\bibinfo  {journal} {Phys.
  Rev. D}\ }\textbf {\bibinfo {volume} {81}},\ \bibinfo {pages} {064002}
  (\bibinfo {year} {2010})},\ \Eprint
  {http://arxiv.org/abs/0912.0399}{arXiv:0912.0399 [hep-th]}\BibitemShut
  {NoStop}%
\bibitem [{\citenamefont {Chen}\ \emph {et~al.}(2009)\citenamefont {Chen},
  \citenamefont {Pi},\ and\ \citenamefont {Tang}}]{Chen:2009vu}%
  \BibitemOpen
  \bibfield  {author} {\bibinfo {author} {\bibfnamefont {B.}~\bibnamefont
  {Chen}}, \bibinfo {author} {\bibfnamefont {S.}~\bibnamefont {Pi}}, \ and\
  \bibinfo {author} {\bibfnamefont {J.-Z.}\ \bibnamefont {Tang}},\ }\bibfield
  {title} {\enquote {\bibinfo {title} {{Power spectra of scalar and tensor
  modes in modified Horava-Lifshitz gravity}},}\ }\href@noop {} {\  (\bibinfo
  {year} {2009})},\ \Eprint {http://arxiv.org/abs/0910.0338}{arXiv:0910.0338
  [hep-th]}\BibitemShut {NoStop}%
\bibitem [{\citenamefont {Janka}(2017)}]{Janka:2017vlw}%
  \BibitemOpen
  \bibfield  {author} {\bibinfo {author} {\bibfnamefont {H.-T.}\ \bibnamefont
  {Janka}},\ }\enquote {\bibinfo {title} {Neutrino emission from supernovae},}\
  in\ \href {\doibase 10.1007/978-3-319-21846-5_4} {\emph {\bibinfo {booktitle}
  {Handbook of Supernovae}}},\ \bibinfo {editor} {edited by\ \bibinfo {editor}
  {\bibfnamefont {A.~W.}\ \bibnamefont {Alsabti}}\ and\ \bibinfo {editor}
  {\bibfnamefont {P.}~\bibnamefont {Murdin}}}\ (\bibinfo  {publisher} {Springer
  International Publishing},\ \bibinfo {address} {Cham},\ \bibinfo {year}
  {2017})\ pp.\ \bibinfo {pages} {1575--1604},\ \Eprint
  {http://arxiv.org/abs/1702.08713}{arXiv:1702.08713 [astro-ph.HE]}\BibitemShut
  {NoStop}%
%%CITATION = ARXIV:1702.08713;%%
\bibitem [{\citenamefont {Piron}(2016)}]{PIRON2016617}%
  \BibitemOpen
  \bibfield  {author} {\bibinfo {author} {\bibfnamefont {F.}~\bibnamefont
  {Piron}},\ }\bibfield  {title} {\enquote {\bibinfo {title} {Gamma-ray bursts
  at high and very high energies},}\ }\href {\doibase
  https://doi.org/10.1016/j.crhy.2016.04.005} {\bibfield  {journal} {\bibinfo
  {journal} {Comptes Rendus Physique}\ }\textbf {\bibinfo {volume} {17}},\
  \bibinfo {pages} {617 } (\bibinfo {year} {2016})},\ \bibinfo {note}
  {gamma-ray astronomy / Astronomie des rayons gamma - Volume 2}\BibitemShut
  {NoStop}%
\bibitem [{\citenamefont {Dunkley}\ \emph {et~al.}(2005)\citenamefont
  {Dunkley}, \citenamefont {Bucher}, \citenamefont {Ferreira}, \citenamefont
  {Moodley},\ and\ \citenamefont {Skordis}}]{Dunkley:2004sv}%
  \BibitemOpen
  \bibfield  {author} {\bibinfo {author} {\bibfnamefont {J.}~\bibnamefont
  {Dunkley}}, \bibinfo {author} {\bibfnamefont {M.}~\bibnamefont {Bucher}},
  \bibinfo {author} {\bibfnamefont {P.~G.}\ \bibnamefont {Ferreira}}, \bibinfo
  {author} {\bibfnamefont {K.}~\bibnamefont {Moodley}}, \ and\ \bibinfo
  {author} {\bibfnamefont {C.}~\bibnamefont {Skordis}},\ }\bibfield  {title}
  {\enquote {\bibinfo {title} {{Fast and reliable mcmc for cosmological
  parameter estimation}},}\ }\href {\doibase 10.1111/j.1365-2966.2004.08464.x}
  {\bibfield  {journal} {\bibinfo  {journal} {Mon. Not. Roy. Astron. Soc.}\
  }\textbf {\bibinfo {volume} {356}},\ \bibinfo {pages} {925} (\bibinfo {year}
  {2005})},\ \Eprint
  {http://arxiv.org/abs/astro-ph/0405462}{arXiv:astro-ph/0405462
  [astro-ph]}\BibitemShut {NoStop}%
%%CITATION = ASTRO-PH/0405462;%%
\bibitem [{\citenamefont {Aghanim}\ \emph {et~al.}(2020)\citenamefont {Aghanim}
  \emph {et~al.}}]{Planck:2018vyg}%
  \BibitemOpen
  \bibfield  {author} {\bibinfo {author} {\bibfnamefont {N.}~\bibnamefont
  {Aghanim}} \emph {et~al.} (\bibinfo {collaboration} {Planck}),\ }\bibfield
  {title} {\enquote {\bibinfo {title} {{Planck 2018 results. VI. Cosmological
  parameters}},}\ }\href {\doibase 10.1051/0004-6361/201833910} {\bibfield
  {journal} {\bibinfo  {journal} {Astron. Astrophys.}\ }\textbf {\bibinfo
  {volume} {641}},\ \bibinfo {pages} {A6} (\bibinfo {year} {2020})},\ \bibinfo
  {note} {[Erratum: Astron.Astrophys. 652, C4 (2021)]},\ \Eprint
  {http://arxiv.org/abs/1807.06209}{arXiv:1807.06209 [astro-ph.CO]}\BibitemShut
  {NoStop}%
\bibitem [{\citenamefont {Carneiro}\ \emph {et~al.}(2019)\citenamefont
  {Carneiro}, \citenamefont {de~Holanda}, \citenamefont {Pigozzo},\ and\
  \citenamefont {Sobreira}}]{Carneiro:2018xwq}%
  \BibitemOpen
  \bibfield  {author} {\bibinfo {author} {\bibfnamefont {S.}~\bibnamefont
  {Carneiro}}, \bibinfo {author} {\bibfnamefont {P.~C.}\ \bibnamefont
  {de~Holanda}}, \bibinfo {author} {\bibfnamefont {C.}~\bibnamefont {Pigozzo}},
  \ and\ \bibinfo {author} {\bibfnamefont {F.}~\bibnamefont {Sobreira}},\
  }\bibfield  {title} {\enquote {\bibinfo {title} {{Is the $H_0$ tension
  suggesting a fourth neutrino generation?}}}\ }\href {\doibase
  10.1103/PhysRevD.100.023505} {\bibfield  {journal} {\bibinfo  {journal}
  {Phys. Rev. D}\ }\textbf {\bibinfo {volume} {100}},\ \bibinfo {pages}
  {023505} (\bibinfo {year} {2019})},\ \Eprint
  {http://arxiv.org/abs/1812.06064}{arXiv:1812.06064 [astro-ph.CO]}\BibitemShut
  {NoStop}%
\bibitem [{\citenamefont {Hagiwara}\ \emph {et~al.}(2002)\citenamefont
  {Hagiwara} \emph {et~al.}}]{Hagiwara:2002fs}%
  \BibitemOpen
  \bibfield  {author} {\bibinfo {author} {\bibfnamefont {K.}~\bibnamefont
  {Hagiwara}} \emph {et~al.} (\bibinfo {collaboration} {Particle Data Group}),\
  }\bibfield  {title} {\enquote {\bibinfo {title} {{Review of particle physics.
  Particle Data Group}},}\ }\href {\doibase 10.1103/PhysRevD.66.010001}
  {\bibfield  {journal} {\bibinfo  {journal} {Phys. Rev.}\ }\textbf {\bibinfo
  {volume} {D66}},\ \bibinfo {pages} {010001} (\bibinfo {year}
  {2002})}\BibitemShut {NoStop}%
%%CITATION = PHRVA,D66,010001;%%
\bibitem [{\citenamefont {Steigman}(2006)}]{Steigman:2005uz}%
  \BibitemOpen
  \bibfield  {author} {\bibinfo {author} {\bibfnamefont {G.}~\bibnamefont
  {Steigman}},\ }\bibfield  {title} {\enquote {\bibinfo {title} {{Primordial
  nucleosynthesis: successes and challenges}},}\ }\href {\doibase
  10.1142/S0218301306004028} {\bibfield  {journal} {\bibinfo  {journal} {Int.
  J. Mod. Phys.}\ }\textbf {\bibinfo {volume} {E15}},\ \bibinfo {pages} {1}
  (\bibinfo {year} {2006})},\ \Eprint
  {http://arxiv.org/abs/astro-ph/0511534}{arXiv:astro-ph/0511534
  [astro-ph]}\BibitemShut {NoStop}%
%%CITATION = ASTRO-PH/0511534;%%
\bibitem [{\citenamefont {Cyburt}\ \emph {et~al.}(2016)\citenamefont {Cyburt},
  \citenamefont {Fields}, \citenamefont {Olive},\ and\ \citenamefont
  {Yeh}}]{Cyburt:2015mya}%
  \BibitemOpen
  \bibfield  {author} {\bibinfo {author} {\bibfnamefont {R.~H.}\ \bibnamefont
  {Cyburt}}, \bibinfo {author} {\bibfnamefont {B.~D.}\ \bibnamefont {Fields}},
  \bibinfo {author} {\bibfnamefont {K.~A.}\ \bibnamefont {Olive}}, \ and\
  \bibinfo {author} {\bibfnamefont {T.-H.}\ \bibnamefont {Yeh}},\ }\bibfield
  {title} {\enquote {\bibinfo {title} {{Big Bang Nucleosynthesis: 2015}},}\
  }\href {\doibase 10.1103/RevModPhys.88.015004} {\bibfield  {journal}
  {\bibinfo  {journal} {Rev. Mod. Phys.}\ }\textbf {\bibinfo {volume} {88}},\
  \bibinfo {pages} {015004} (\bibinfo {year} {2016})},\ \Eprint
  {http://arxiv.org/abs/1505.01076}{arXiv:1505.01076 [astro-ph.CO]}\BibitemShut
  {NoStop}%
%%CITATION = ARXIV:1505.01076;%%
\bibitem [{\citenamefont {Di~Valentino}\ \emph {et~al.}(2019)\citenamefont
  {Di~Valentino}, \citenamefont {Melchiorri},\ and\ \citenamefont
  {Silk}}]{DiValentino:2019qzk}%
  \BibitemOpen
  \bibfield  {author} {\bibinfo {author} {\bibfnamefont {E.}~\bibnamefont
  {Di~Valentino}}, \bibinfo {author} {\bibfnamefont {A.}~\bibnamefont
  {Melchiorri}}, \ and\ \bibinfo {author} {\bibfnamefont {J.}~\bibnamefont
  {Silk}},\ }\bibfield  {title} {\enquote {\bibinfo {title} {{Planck evidence
  for a closed Universe and a possible crisis for cosmology}},}\ }\href
  {\doibase 10.1038/s41550-019-0906-9} {\bibfield  {journal} {\bibinfo
  {journal} {Nature Astron.}\ }\textbf {\bibinfo {volume} {4}},\ \bibinfo
  {pages} {196} (\bibinfo {year} {2019})},\ \Eprint
  {http://arxiv.org/abs/1911.02087}{arXiv:1911.02087 [astro-ph.CO]}\BibitemShut
  {NoStop}%
\bibitem [{\citenamefont {Abbott}\ \emph
  {et~al.}(2018{\natexlab{a}})\citenamefont {Abbott} \emph
  {et~al.}}]{LIGOScientific:2017ous}%
  \BibitemOpen
  \bibfield  {author} {\bibinfo {author} {\bibfnamefont {B.~P.}\ \bibnamefont
  {Abbott}} \emph {et~al.} (\bibinfo {collaboration} {LIGO Scientific,
  Virgo}),\ }\bibfield  {title} {\enquote {\bibinfo {title} {{First search for
  nontensorial gravitational waves from known pulsars}},}\ }\href {\doibase
  10.1103/PhysRevLett.120.031104} {\bibfield  {journal} {\bibinfo  {journal}
  {Phys. Rev. Lett.}\ }\textbf {\bibinfo {volume} {120}},\ \bibinfo {pages}
  {031104} (\bibinfo {year} {2018}{\natexlab{a}})},\ \Eprint
  {http://arxiv.org/abs/1709.09203}{arXiv:1709.09203 [gr-qc]}\BibitemShut
  {NoStop}%
\bibitem [{\citenamefont {Abbott}\ \emph
  {et~al.}(2018{\natexlab{b}})\citenamefont {Abbott} \emph
  {et~al.}}]{LIGOScientific:2018czr}%
  \BibitemOpen
  \bibfield  {author} {\bibinfo {author} {\bibfnamefont {B.~P.}\ \bibnamefont
  {Abbott}} \emph {et~al.} (\bibinfo {collaboration} {LIGO Scientific,
  Virgo}),\ }\bibfield  {title} {\enquote {\bibinfo {title} {{Search for
  Tensor, Vector, and Scalar Polarizations in the Stochastic Gravitational-Wave
  Background}},}\ }\href {\doibase 10.1103/PhysRevLett.120.201102} {\bibfield
  {journal} {\bibinfo  {journal} {Phys. Rev. Lett.}\ }\textbf {\bibinfo
  {volume} {120}},\ \bibinfo {pages} {201102} (\bibinfo {year}
  {2018}{\natexlab{b}})},\ \Eprint
  {http://arxiv.org/abs/1802.10194}{arXiv:1802.10194 [gr-qc]}\BibitemShut
  {NoStop}%
\bibitem [{\citenamefont {Liberati}\ \emph {et~al.}(2012)\citenamefont
  {Liberati}, \citenamefont {Maccione},\ and\ \citenamefont
  {Sotiriou}}]{PhysRevLett.109.151602}%
  \BibitemOpen
  \bibfield  {author} {\bibinfo {author} {\bibfnamefont {S.}~\bibnamefont
  {Liberati}}, \bibinfo {author} {\bibfnamefont {L.}~\bibnamefont {Maccione}},
  \ and\ \bibinfo {author} {\bibfnamefont {T.~P.}\ \bibnamefont {Sotiriou}},\
  }\bibfield  {title} {\enquote {\bibinfo {title} {Scale hierarchy in
  ho\ifmmode \check{r}\else \v{r}\fi{}ava-lifshitz gravity: Strong constraint
  from synchrotron radiation in the crab nebula},}\ }\href {\doibase
  10.1103/PhysRevLett.109.151602} {\bibfield  {journal} {\bibinfo  {journal}
  {Phys. Rev. Lett.}\ }\textbf {\bibinfo {volume} {109}},\ \bibinfo {pages}
  {151602} (\bibinfo {year} {2012})}\BibitemShut {NoStop}%
\bibitem [{\citenamefont {Kostelecky}\ and\ \citenamefont
  {Russell}(2011)}]{Kostelecky:2008ts}%
  \BibitemOpen
  \bibfield  {author} {\bibinfo {author} {\bibfnamefont {V.~A.}\ \bibnamefont
  {Kostelecky}}\ and\ \bibinfo {author} {\bibfnamefont {N.}~\bibnamefont
  {Russell}},\ }\bibfield  {title} {\enquote {\bibinfo {title} {{Data Tables
  for Lorentz and CPT Violation}},}\ }\href {\doibase 10.1103/RevModPhys.83.11}
  {\bibfield  {journal} {\bibinfo  {journal} {Rev. Mod. Phys.}\ }\textbf
  {\bibinfo {volume} {83}},\ \bibinfo {pages} {11} (\bibinfo {year} {2011})},\
  \Eprint {http://arxiv.org/abs/0801.0287}{arXiv:0801.0287
  [hep-ph]}\BibitemShut {NoStop}%
\bibitem [{\citenamefont {Kostelecky}\ and\ \citenamefont
  {Tasson}(2011)}]{Kostelecky:2010ze}%
  \BibitemOpen
  \bibfield  {author} {\bibinfo {author} {\bibfnamefont {A.~V.}\ \bibnamefont
  {Kostelecky}}\ and\ \bibinfo {author} {\bibfnamefont {J.~D.}\ \bibnamefont
  {Tasson}},\ }\bibfield  {title} {\enquote {\bibinfo {title} {{Matter-gravity
  couplings and Lorentz violation}},}\ }\href {\doibase
  10.1103/PhysRevD.83.016013} {\bibfield  {journal} {\bibinfo  {journal} {Phys.
  Rev. D}\ }\textbf {\bibinfo {volume} {83}},\ \bibinfo {pages} {016013}
  (\bibinfo {year} {2011})},\ \Eprint
  {http://arxiv.org/abs/1006.4106}{arXiv:1006.4106 [gr-qc]}\BibitemShut
  {NoStop}%
\bibitem [{\citenamefont {Colombo}\ \emph
  {et~al.}(2015{\natexlab{b}})\citenamefont {Colombo}, \citenamefont
  {G\"umr\"uk\ifmmode \mbox{\c{c}}\else \c{c}\fi{}\"uo\ifmmode~\breve{g}\else
  \u{g}\fi{}lu},\ and\ \citenamefont {Sotiriou}}]{PhysRevD.91.044021}%
  \BibitemOpen
  \bibfield  {author} {\bibinfo {author} {\bibfnamefont {M.}~\bibnamefont
  {Colombo}}, \bibinfo {author} {\bibfnamefont {A.~E.}\ \bibnamefont
  {G\"umr\"uk\ifmmode \mbox{\c{c}}\else \c{c}\fi{}\"uo\ifmmode~\breve{g}\else
  \u{g}\fi{}lu}}, \ and\ \bibinfo {author} {\bibfnamefont {T.~P.}\ \bibnamefont
  {Sotiriou}},\ }\bibfield  {title} {\enquote {\bibinfo {title} {Ho\ifmmode
  \check{r}\else \v{r}\fi{}ava gravity with mixed derivative terms},}\ }\href
  {\doibase 10.1103/PhysRevD.91.044021} {\bibfield  {journal} {\bibinfo
  {journal} {Phys. Rev. D}\ }\textbf {\bibinfo {volume} {91}},\ \bibinfo
  {pages} {044021} (\bibinfo {year} {2015}{\natexlab{b}})}\BibitemShut
  {NoStop}%
\bibitem [{\citenamefont {Colombo}\ \emph
  {et~al.}(2015{\natexlab{c}})\citenamefont {Colombo}, \citenamefont
  {G\"umr\"uk\c{c}\"uo\u{g}lu},\ and\ \citenamefont
  {Sotiriou}}]{PhysRevD.92.064037}%
  \BibitemOpen
  \bibfield  {author} {\bibinfo {author} {\bibfnamefont {M.}~\bibnamefont
  {Colombo}}, \bibinfo {author} {\bibfnamefont {A.~E.}\ \bibnamefont
  {G\"umr\"uk\c{c}\"uo\u{g}lu}}, \ and\ \bibinfo {author} {\bibfnamefont
  {T.~P.}\ \bibnamefont {Sotiriou}},\ }\bibfield  {title} {\enquote {\bibinfo
  {title} {{Ho\v{r}ava gravity with mixed derivative terms: Power counting
  renormalizability with lower order dispersions}},}\ }\href {\doibase
  10.1103/PhysRevD.92.064037} {\bibfield  {journal} {\bibinfo  {journal} {Phys.
  Rev. D}\ }\textbf {\bibinfo {volume} {92}},\ \bibinfo {pages} {064037}
  (\bibinfo {year} {2015}{\natexlab{c}})},\ \Eprint
  {http://arxiv.org/abs/1503.07544}{arXiv:1503.07544 [hep-th]}\BibitemShut
  {NoStop}%
\bibitem [{\citenamefont {Coates}\ \emph {et~al.}(2016)\citenamefont {Coates},
  \citenamefont {Colombo}, \citenamefont {G\"umr\"uk\ifmmode \mbox{\c{c}}\else
  \c{c}\fi{}\"uo\ifmmode~\breve{g}\else \u{g}\fi{}lu},\ and\ \citenamefont
  {Sotiriou}}]{PhysRevD.94.084014}%
  \BibitemOpen
  \bibfield  {author} {\bibinfo {author} {\bibfnamefont {A.}~\bibnamefont
  {Coates}}, \bibinfo {author} {\bibfnamefont {M.}~\bibnamefont {Colombo}},
  \bibinfo {author} {\bibfnamefont {A.~E.}\ \bibnamefont {G\"umr\"uk\ifmmode
  \mbox{\c{c}}\else \c{c}\fi{}\"uo\ifmmode~\breve{g}\else \u{g}\fi{}lu}}, \
  and\ \bibinfo {author} {\bibfnamefont {T.~P.}\ \bibnamefont {Sotiriou}},\
  }\bibfield  {title} {\enquote {\bibinfo {title} {Uninvited guest in mixed
  derivative ho\ifmmode \check{r}\else \v{r}\fi{}ava gravity},}\ }\href
  {\doibase 10.1103/PhysRevD.94.084014} {\bibfield  {journal} {\bibinfo
  {journal} {Phys. Rev. D}\ }\textbf {\bibinfo {volume} {94}},\ \bibinfo
  {pages} {084014} (\bibinfo {year} {2016})}\BibitemShut {NoStop}%
\bibitem [{\citenamefont {Kluso\ifmmode~\check{n}\else
  \v{n}\fi{}}(2016)}]{PhysRevD.94.104043}%
  \BibitemOpen
  \bibfield  {author} {\bibinfo {author} {\bibfnamefont {J.}~\bibnamefont
  {Kluso\ifmmode~\check{n}\else \v{n}\fi{}}},\ }\bibfield  {title} {\enquote
  {\bibinfo {title} {Hamiltonian analysis of mixed derivative ho\ifmmode
  \check{r}\else \v{r}\fi{}ava-lifshitz gravity},}\ }\href {\doibase
  10.1103/PhysRevD.94.104043} {\bibfield  {journal} {\bibinfo  {journal} {Phys.
  Rev. D}\ }\textbf {\bibinfo {volume} {94}},\ \bibinfo {pages} {104043}
  (\bibinfo {year} {2016})}\BibitemShut {NoStop}%
\bibitem [{\citenamefont {da~Silva}(2011)}]{daSilva:2010bm}%
  \BibitemOpen
  \bibfield  {author} {\bibinfo {author} {\bibfnamefont {A.~M.}\ \bibnamefont
  {da~Silva}},\ }\bibfield  {title} {\enquote {\bibinfo {title} {{An
  Alternative Approach for General Covariant Horava-Lifshitz Gravity and Matter
  Coupling}},}\ }\href {\doibase 10.1088/0264-9381/28/5/055011} {\bibfield
  {journal} {\bibinfo  {journal} {Class. Quant. Grav.}\ }\textbf {\bibinfo
  {volume} {28}},\ \bibinfo {pages} {055011} (\bibinfo {year} {2011})},\
  \Eprint {http://arxiv.org/abs/1009.4885}{arXiv:1009.4885
  [hep-th]}\BibitemShut {NoStop}%
\bibitem [{\citenamefont {Bellorin}\ and\ \citenamefont
  {Restuccia}(2012)}]{Bellorin:2010je}%
  \BibitemOpen
  \bibfield  {author} {\bibinfo {author} {\bibfnamefont {J.}~\bibnamefont
  {Bellorin}}\ and\ \bibinfo {author} {\bibfnamefont {A.}~\bibnamefont
  {Restuccia}},\ }\bibfield  {title} {\enquote {\bibinfo {title} {{On the
  consistency of the Horava Theory}},}\ }\href {\doibase
  10.1142/S0218271812500290} {\bibfield  {journal} {\bibinfo  {journal} {Int.
  J. Mod. Phys. D}\ }\textbf {\bibinfo {volume} {21}},\ \bibinfo {pages}
  {1250029} (\bibinfo {year} {2012})},\ \Eprint
  {http://arxiv.org/abs/1004.0055}{arXiv:1004.0055 [hep-th]}\BibitemShut
  {NoStop}%
\bibitem [{\citenamefont {Loll}\ and\ \citenamefont
  {Pires}(2014)}]{Loll:2014xja}%
  \BibitemOpen
  \bibfield  {author} {\bibinfo {author} {\bibfnamefont {R.}~\bibnamefont
  {Loll}}\ and\ \bibinfo {author} {\bibfnamefont {L.}~\bibnamefont {Pires}},\
  }\bibfield  {title} {\enquote {\bibinfo {title} {{Role of the extra coupling
  in the kinetic term in Ho\v{r}ava-Lifshitz gravity}},}\ }\href {\doibase
  10.1103/PhysRevD.90.124050} {\bibfield  {journal} {\bibinfo  {journal} {Phys.
  Rev. D}\ }\textbf {\bibinfo {volume} {90}},\ \bibinfo {pages} {124050}
  (\bibinfo {year} {2014})},\ \Eprint
  {http://arxiv.org/abs/1407.1259}{arXiv:1407.1259 [hep-th]}\BibitemShut
  {NoStop}%
\bibitem [{\citenamefont {Hu}\ and\ \citenamefont
  {Sugiyama}(1996)}]{Hu:1995en}%
  \BibitemOpen
  \bibfield  {author} {\bibinfo {author} {\bibfnamefont {W.}~\bibnamefont
  {Hu}}\ and\ \bibinfo {author} {\bibfnamefont {N.}~\bibnamefont {Sugiyama}},\
  }\bibfield  {title} {\enquote {\bibinfo {title} {{Small scale cosmological
  perturbations: An Analytic approach}},}\ }\href {\doibase 10.1086/177989}
  {\bibfield  {journal} {\bibinfo  {journal} {Astrophys. J.}\ }\textbf
  {\bibinfo {volume} {471}},\ \bibinfo {pages} {542} (\bibinfo {year}
  {1996})},\ \Eprint
  {http://arxiv.org/abs/astro-ph/9510117}{arXiv:astro-ph/9510117}\BibitemShut
  {NoStop}%
\bibitem [{\citenamefont {Zhai}\ \emph {et~al.}(2020)\citenamefont {Zhai},
  \citenamefont {Park}, \citenamefont {Wang},\ and\ \citenamefont
  {Ratra}}]{Zhai:2019nad}%
  \BibitemOpen
  \bibfield  {author} {\bibinfo {author} {\bibfnamefont {Z.}~\bibnamefont
  {Zhai}}, \bibinfo {author} {\bibfnamefont {C.-G.}\ \bibnamefont {Park}},
  \bibinfo {author} {\bibfnamefont {Y.}~\bibnamefont {Wang}}, \ and\ \bibinfo
  {author} {\bibfnamefont {B.}~\bibnamefont {Ratra}},\ }\bibfield  {title}
  {\enquote {\bibinfo {title} {{CMB distance priors revisited: effects of dark
  energy dynamics, spatial curvature, primordial power spectrum, and neutrino
  parameters}},}\ }\href {\doibase 10.1088/1475-7516/2020/07/009} {\bibfield
  {journal} {\bibinfo  {journal} {JCAP}\ }\textbf {\bibinfo {volume} {07}},\
  \bibinfo {pages} {009} (\bibinfo {year} {2020})},\ \Eprint
  {http://arxiv.org/abs/1912.04921}{arXiv:1912.04921 [astro-ph.CO]}\BibitemShut
  {NoStop}%
\bibitem [{\citenamefont {Scolnic}\ \emph {et~al.}(2022)\citenamefont {Scolnic}
  \emph {et~al.}}]{Scolnic:2021amr}%
  \BibitemOpen
  \bibfield  {author} {\bibinfo {author} {\bibfnamefont {D.}~\bibnamefont
  {Scolnic}} \emph {et~al.},\ }\bibfield  {title} {\enquote {\bibinfo {title}
  {{The Pantheon+ Analysis: The Full Data Set and Light-curve Release}},}\
  }\href {\doibase 10.3847/1538-4357/ac8b7a} {\bibfield  {journal} {\bibinfo
  {journal} {Astrophys. J.}\ }\textbf {\bibinfo {volume} {938}},\ \bibinfo
  {pages} {113} (\bibinfo {year} {2022})},\ \Eprint
  {http://arxiv.org/abs/2112.03863}{arXiv:2112.03863 [astro-ph.CO]}\BibitemShut
  {NoStop}%
\bibitem [{\citenamefont {Brout}\ \emph {et~al.}(2022)\citenamefont {Brout}
  \emph {et~al.}}]{Brout:2022vxf}%
  \BibitemOpen
  \bibfield  {author} {\bibinfo {author} {\bibfnamefont {D.}~\bibnamefont
  {Brout}} \emph {et~al.},\ }\bibfield  {title} {\enquote {\bibinfo {title}
  {{The Pantheon+ Analysis: Cosmological Constraints}},}\ }\href {\doibase
  10.3847/1538-4357/ac8e04} {\bibfield  {journal} {\bibinfo  {journal}
  {Astrophys. J.}\ }\textbf {\bibinfo {volume} {938}},\ \bibinfo {pages} {110}
  (\bibinfo {year} {2022})},\ \Eprint
  {http://arxiv.org/abs/2202.04077}{arXiv:2202.04077 [astro-ph.CO]}\BibitemShut
  {NoStop}%
\bibitem [{\citenamefont {Liu}\ and\ \citenamefont {Wei}(2015)}]{Liu:2014vda}%
  \BibitemOpen
  \bibfield  {author} {\bibinfo {author} {\bibfnamefont {J.}~\bibnamefont
  {Liu}}\ and\ \bibinfo {author} {\bibfnamefont {H.}~\bibnamefont {Wei}},\
  }\bibfield  {title} {\enquote {\bibinfo {title} {{Cosmological models and
  gamma-ray bursts calibrated by using Pade method}},}\ }\href {\doibase
  10.1007/s10714-015-1986-1} {\bibfield  {journal} {\bibinfo  {journal} {Gen.
  Rel. Grav.}\ }\textbf {\bibinfo {volume} {47}},\ \bibinfo {pages} {141}
  (\bibinfo {year} {2015})},\ \Eprint
  {http://arxiv.org/abs/1410.3960}{arXiv:1410.3960 [astro-ph.CO]}\BibitemShut
  {NoStop}%
%%CITATION = ARXIV:1410.3960;%%
\bibitem [{\citenamefont {Ghirlanda}\ \emph {et~al.}(2006)\citenamefont
  {Ghirlanda}, \citenamefont {Ghisellini},\ and\ \citenamefont
  {Firmani}}]{Ghirlanda:2006ax}%
  \BibitemOpen
  \bibfield  {author} {\bibinfo {author} {\bibfnamefont {G.}~\bibnamefont
  {Ghirlanda}}, \bibinfo {author} {\bibfnamefont {G.}~\bibnamefont
  {Ghisellini}}, \ and\ \bibinfo {author} {\bibfnamefont {C.}~\bibnamefont
  {Firmani}},\ }\bibfield  {title} {\enquote {\bibinfo {title} {{Gamma Ray
  Bursts as standard candles to constrain the cosmological parameters}},}\
  }\href {\doibase 10.1088/1367-2630/8/7/123} {\bibfield  {journal} {\bibinfo
  {journal} {New J. Phys.}\ }\textbf {\bibinfo {volume} {8}},\ \bibinfo {pages}
  {123} (\bibinfo {year} {2006})},\ \Eprint
  {http://arxiv.org/abs/astro-ph/0610248}{arXiv:astro-ph/0610248
  [astro-ph]}\BibitemShut {NoStop}%
%%CITATION = ASTRO-PH/0610248;%%
\bibitem [{\citenamefont {Jimenez}\ and\ \citenamefont
  {Loeb}(2002)}]{Jimenez:2001gg}%
  \BibitemOpen
  \bibfield  {author} {\bibinfo {author} {\bibfnamefont {R.}~\bibnamefont
  {Jimenez}}\ and\ \bibinfo {author} {\bibfnamefont {A.}~\bibnamefont {Loeb}},\
  }\bibfield  {title} {\enquote {\bibinfo {title} {{Constraining cosmological
  parameters based on relative galaxy ages}},}\ }\href {\doibase
  10.1086/340549} {\bibfield  {journal} {\bibinfo  {journal} {Astrophys. J.}\
  }\textbf {\bibinfo {volume} {573}},\ \bibinfo {pages} {37} (\bibinfo {year}
  {2002})},\ \Eprint
  {http://arxiv.org/abs/astro-ph/0106145}{arXiv:astro-ph/0106145}\BibitemShut
  {NoStop}%
\bibitem [{\citenamefont {Moresco}\ \emph {et~al.}(2012)\citenamefont
  {Moresco}, \citenamefont {Verde}, \citenamefont {Pozzetti}, \citenamefont
  {Jimenez},\ and\ \citenamefont {Cimatti}}]{Moresco:2012by}%
  \BibitemOpen
  \bibfield  {author} {\bibinfo {author} {\bibfnamefont {M.}~\bibnamefont
  {Moresco}}, \bibinfo {author} {\bibfnamefont {L.}~\bibnamefont {Verde}},
  \bibinfo {author} {\bibfnamefont {L.}~\bibnamefont {Pozzetti}}, \bibinfo
  {author} {\bibfnamefont {R.}~\bibnamefont {Jimenez}}, \ and\ \bibinfo
  {author} {\bibfnamefont {A.}~\bibnamefont {Cimatti}},\ }\bibfield  {title}
  {\enquote {\bibinfo {title} {{New constraints on cosmological parameters and
  neutrino properties using the expansion rate of the Universe to
  z\textasciitilde{}1.75}},}\ }\href {\doibase 10.1088/1475-7516/2012/07/053}
  {\bibfield  {journal} {\bibinfo  {journal} {JCAP}\ }\textbf {\bibinfo
  {volume} {07}},\ \bibinfo {pages} {053} (\bibinfo {year} {2012})},\ \Eprint
  {http://arxiv.org/abs/1201.6658}{arXiv:1201.6658 [astro-ph.CO]}\BibitemShut
  {NoStop}%
\bibitem [{\citenamefont {Moresco}(2015)}]{Moresco:2015cya}%
  \BibitemOpen
  \bibfield  {author} {\bibinfo {author} {\bibfnamefont {M.}~\bibnamefont
  {Moresco}},\ }\bibfield  {title} {\enquote {\bibinfo {title} {{Raising the
  bar: new constraints on the Hubble parameter with cosmic chronometers at z
  \ensuremath{\sim} 2}},}\ }\href {\doibase 10.1093/mnrasl/slv037} {\bibfield
  {journal} {\bibinfo  {journal} {Mon. Not. Roy. Astron. Soc.}\ }\textbf
  {\bibinfo {volume} {450}},\ \bibinfo {pages} {L16} (\bibinfo {year}
  {2015})},\ \Eprint {http://arxiv.org/abs/1503.01116}{arXiv:1503.01116
  [astro-ph.CO]}\BibitemShut {NoStop}%
\bibitem [{\citenamefont {G\'omez-Valent}\ and\ \citenamefont
  {Amendola}(2019)}]{Gomez-Valent:2019lny}%
  \BibitemOpen
  \bibfield  {author} {\bibinfo {author} {\bibfnamefont {A.}~\bibnamefont
  {G\'omez-Valent}}\ and\ \bibinfo {author} {\bibfnamefont {L.}~\bibnamefont
  {Amendola}},\ }\bibfield  {title} {\enquote {\bibinfo {title} {{$H_0$ from
  cosmic chronometers and Type Ia supernovae, with Gaussian processes and the
  weighted polynomial regression method}},}\ }in\ \href@noop {} {\emph
  {\bibinfo {booktitle} {{15th Marcel Grossmann Meeting on Recent Developments
  in Theoretical and Experimental General Relativity, Astrophysics, and
  Relativistic Field Theories}}}}\ (\bibinfo {year} {2019})\ \Eprint
  {http://arxiv.org/abs/1905.04052}{arXiv:1905.04052 [astro-ph.CO]}\BibitemShut
  {NoStop}%
\bibitem [{\citenamefont {Moresco}\ \emph {et~al.}(2022)\citenamefont {Moresco}
  \emph {et~al.}}]{Moresco:2022phi}%
  \BibitemOpen
  \bibfield  {author} {\bibinfo {author} {\bibfnamefont {M.}~\bibnamefont
  {Moresco}} \emph {et~al.},\ }\bibfield  {title} {\enquote {\bibinfo {title}
  {{Unveiling the Universe with emerging cosmological probes}},}\ }\href
  {\doibase 10.1007/s41114-022-00040-z} {\bibfield  {journal} {\bibinfo
  {journal} {Living Rev. Rel.}\ }\textbf {\bibinfo {volume} {25}},\ \bibinfo
  {pages} {6} (\bibinfo {year} {2022})},\ \Eprint
  {http://arxiv.org/abs/2201.07241}{arXiv:2201.07241 [astro-ph.CO]}\BibitemShut
  {NoStop}%
\bibitem [{\citenamefont {Moresco}\ \emph {et~al.}(2020)\citenamefont
  {Moresco}, \citenamefont {Jimenez}, \citenamefont {Verde}, \citenamefont
  {Cimatti},\ and\ \citenamefont {Pozzetti}}]{Moresco:2020fbm}%
  \BibitemOpen
  \bibfield  {author} {\bibinfo {author} {\bibfnamefont {M.}~\bibnamefont
  {Moresco}}, \bibinfo {author} {\bibfnamefont {R.}~\bibnamefont {Jimenez}},
  \bibinfo {author} {\bibfnamefont {L.}~\bibnamefont {Verde}}, \bibinfo
  {author} {\bibfnamefont {A.}~\bibnamefont {Cimatti}}, \ and\ \bibinfo
  {author} {\bibfnamefont {L.}~\bibnamefont {Pozzetti}},\ }\bibfield  {title}
  {\enquote {\bibinfo {title} {{Setting the Stage for Cosmic Chronometers. II.
  Impact of Stellar Population Synthesis Models Systematics and Full Covariance
  Matrix}},}\ }\href {\doibase 10.3847/1538-4357/ab9eb0} {\bibfield  {journal}
  {\bibinfo  {journal} {Astrophys. J.}\ }\textbf {\bibinfo {volume} {898}},\
  \bibinfo {pages} {82} (\bibinfo {year} {2020})},\ \Eprint
  {http://arxiv.org/abs/2003.07362}{arXiv:2003.07362 [astro-ph.GA]}\BibitemShut
  {NoStop}%
\bibitem [{\citenamefont {{Blake}}\ \emph {et~al.}(2012)\citenamefont
  {{Blake}}, \citenamefont {{Brough}}, \citenamefont {{Colless}}, \citenamefont
  {{Contreras}}, \citenamefont {{Couch}}, \citenamefont {{Croom}},
  \citenamefont {{Croton}}, \citenamefont {{Davis}}, \citenamefont
  {{Drinkwater}}, \citenamefont {{Forster}}, \citenamefont {{Gilbank}},
  \citenamefont {{Gladders}}, \citenamefont {{Glazebrook}}, \citenamefont
  {{Jelliffe}}, \citenamefont {{Jurek}}, \citenamefont {{Li}}, \citenamefont
  {{Madore}}, \citenamefont {{Martin}}, \citenamefont {{Pimbblet}},
  \citenamefont {{Poole}}, \citenamefont {{Pracy}}, \citenamefont {{Sharp}},
  \citenamefont {{Wisnioski}}, \citenamefont {{Woods}}, \citenamefont
  {{Wyder}},\ and\ \citenamefont {{Yee}}}]{2012MNRAS.425..405B}%
  \BibitemOpen
  \bibfield  {author} {\bibinfo {author} {\bibfnamefont {C.}~\bibnamefont
  {{Blake}}}, \bibinfo {author} {\bibfnamefont {S.}~\bibnamefont {{Brough}}},
  \bibinfo {author} {\bibfnamefont {M.}~\bibnamefont {{Colless}}}, \bibinfo
  {author} {\bibfnamefont {C.}~\bibnamefont {{Contreras}}}, \bibinfo {author}
  {\bibfnamefont {W.}~\bibnamefont {{Couch}}}, \bibinfo {author} {\bibfnamefont
  {S.}~\bibnamefont {{Croom}}}, \bibinfo {author} {\bibfnamefont
  {D.}~\bibnamefont {{Croton}}}, \bibinfo {author} {\bibfnamefont {T.~M.}\
  \bibnamefont {{Davis}}}, \bibinfo {author} {\bibfnamefont {M.~J.}\
  \bibnamefont {{Drinkwater}}}, \bibinfo {author} {\bibfnamefont
  {K.}~\bibnamefont {{Forster}}}, \bibinfo {author} {\bibfnamefont
  {D.}~\bibnamefont {{Gilbank}}}, \bibinfo {author} {\bibfnamefont
  {M.}~\bibnamefont {{Gladders}}}, \bibinfo {author} {\bibfnamefont
  {K.}~\bibnamefont {{Glazebrook}}}, \bibinfo {author} {\bibfnamefont
  {B.}~\bibnamefont {{Jelliffe}}}, \bibinfo {author} {\bibfnamefont {R.~J.}\
  \bibnamefont {{Jurek}}}, \bibinfo {author} {\bibfnamefont {I.~h.}\
  \bibnamefont {{Li}}}, \bibinfo {author} {\bibfnamefont {B.}~\bibnamefont
  {{Madore}}}, \bibinfo {author} {\bibfnamefont {D.~C.}\ \bibnamefont
  {{Martin}}}, \bibinfo {author} {\bibfnamefont {K.}~\bibnamefont
  {{Pimbblet}}}, \bibinfo {author} {\bibfnamefont {G.~B.}\ \bibnamefont
  {{Poole}}}, \bibinfo {author} {\bibfnamefont {M.}~\bibnamefont {{Pracy}}},
  \bibinfo {author} {\bibfnamefont {R.}~\bibnamefont {{Sharp}}}, \bibinfo
  {author} {\bibfnamefont {E.}~\bibnamefont {{Wisnioski}}}, \bibinfo {author}
  {\bibfnamefont {D.}~\bibnamefont {{Woods}}}, \bibinfo {author} {\bibfnamefont
  {T.~K.}\ \bibnamefont {{Wyder}}}, \ and\ \bibinfo {author} {\bibfnamefont
  {H.~K.~C.}\ \bibnamefont {{Yee}}},\ }\bibfield  {title} {\enquote {\bibinfo
  {title} {{The WiggleZ Dark Energy Survey: joint measurements of the expansion
  and growth history at z < 1}},}\ }\href {\doibase
  10.1111/j.1365-2966.2012.21473.x} {\bibfield  {journal} {\bibinfo  {journal}
  {MNRAS}\ }\textbf {\bibinfo {volume} {425}},\ \bibinfo {pages} {405}
  (\bibinfo {year} {2012})},\ \Eprint
  {http://arxiv.org/abs/1204.3674}{arXiv:1204.3674 [astro-ph.CO]}\BibitemShut
  {NoStop}%
\bibitem [{\citenamefont {Eisenstein}\ and\ \citenamefont
  {Hu}(1998)}]{Eisenstein:1997ik}%
  \BibitemOpen
  \bibfield  {author} {\bibinfo {author} {\bibfnamefont {D.~J.}\ \bibnamefont
  {Eisenstein}}\ and\ \bibinfo {author} {\bibfnamefont {W.}~\bibnamefont
  {Hu}},\ }\bibfield  {title} {\enquote {\bibinfo {title} {{Baryonic features
  in the matter transfer function}},}\ }\href {\doibase 10.1086/305424}
  {\bibfield  {journal} {\bibinfo  {journal} {Astrophys. J.}\ }\textbf
  {\bibinfo {volume} {496}},\ \bibinfo {pages} {605} (\bibinfo {year}
  {1998})},\ \Eprint
  {http://arxiv.org/abs/astro-ph/9709112}{arXiv:astro-ph/9709112}\BibitemShut
  {NoStop}%
\bibitem [{\citenamefont {Bautista}\ \emph {et~al.}(2020)\citenamefont
  {Bautista} \emph {et~al.}}]{Bautista:2020ahg}%
  \BibitemOpen
  \bibfield  {author} {\bibinfo {author} {\bibfnamefont {J.~E.}\ \bibnamefont
  {Bautista}} \emph {et~al.},\ }\bibfield  {title} {\enquote {\bibinfo {title}
  {{The Completed SDSS-IV extended Baryon Oscillation Spectroscopic Survey:
  measurement of the BAO and growth rate of structure of the luminous red
  galaxy sample from the anisotropic correlation function between redshifts 0.6
  and 1}},}\ }\href {\doibase 10.1093/mnras/staa2800} {\bibfield  {journal}
  {\bibinfo  {journal} {Mon. Not. Roy. Astron. Soc.}\ }\textbf {\bibinfo
  {volume} {500}},\ \bibinfo {pages} {736} (\bibinfo {year} {2020})},\ \Eprint
  {http://arxiv.org/abs/2007.08993}{arXiv:2007.08993 [astro-ph.CO]}\BibitemShut
  {NoStop}%
\bibitem [{\citenamefont {Alam}\ \emph {et~al.}(2017)\citenamefont {Alam} \emph
  {et~al.}}]{BOSS:2016wmc}%
  \BibitemOpen
  \bibfield  {author} {\bibinfo {author} {\bibfnamefont {S.}~\bibnamefont
  {Alam}} \emph {et~al.} (\bibinfo {collaboration} {BOSS}),\ }\bibfield
  {title} {\enquote {\bibinfo {title} {{The clustering of galaxies in the
  completed SDSS-III Baryon Oscillation Spectroscopic Survey: cosmological
  analysis of the DR12 galaxy sample}},}\ }\href {\doibase
  10.1093/mnras/stx721} {\bibfield  {journal} {\bibinfo  {journal} {Mon. Not.
  Roy. Astron. Soc.}\ }\textbf {\bibinfo {volume} {470}},\ \bibinfo {pages}
  {2617} (\bibinfo {year} {2017})},\ \Eprint
  {http://arxiv.org/abs/1607.03155}{arXiv:1607.03155 [astro-ph.CO]}\BibitemShut
  {NoStop}%
\bibitem [{\citenamefont {Zhao}\ \emph {et~al.}(2019)\citenamefont {Zhao} \emph
  {et~al.}}]{Zhao:2018gvb}%
  \BibitemOpen
  \bibfield  {author} {\bibinfo {author} {\bibfnamefont {G.-B.}\ \bibnamefont
  {Zhao}} \emph {et~al.},\ }\bibfield  {title} {\enquote {\bibinfo {title}
  {{The clustering of the SDSS-IV extended Baryon Oscillation Spectroscopic
  Survey DR14 quasar sample: a tomographic measurement of cosmic structure
  growth and expansion rate based on optimal redshift weights}},}\ }\href
  {\doibase 10.1093/mnras/sty2845} {\bibfield  {journal} {\bibinfo  {journal}
  {Mon. Not. Roy. Astron. Soc.}\ }\textbf {\bibinfo {volume} {482}},\ \bibinfo
  {pages} {3497} (\bibinfo {year} {2019})},\ \Eprint
  {http://arxiv.org/abs/1801.03043}{arXiv:1801.03043 [astro-ph.CO]}\BibitemShut
  {NoStop}%
\bibitem [{\citenamefont {Hou}\ \emph {et~al.}(2020)\citenamefont {Hou} \emph
  {et~al.}}]{Hou:2020rse}%
  \BibitemOpen
  \bibfield  {author} {\bibinfo {author} {\bibfnamefont {J.}~\bibnamefont
  {Hou}} \emph {et~al.},\ }\bibfield  {title} {\enquote {\bibinfo {title} {{The
  Completed SDSS-IV extended Baryon Oscillation Spectroscopic Survey: BAO and
  RSD measurements from anisotropic clustering analysis of the Quasar Sample in
  configuration space between redshift 0.8 and 2.2}},}\ }\href {\doibase
  10.1093/mnras/staa3234} {\bibfield  {journal} {\bibinfo  {journal} {Mon. Not.
  Roy. Astron. Soc.}\ }\textbf {\bibinfo {volume} {500}},\ \bibinfo {pages}
  {1201} (\bibinfo {year} {2020})},\ \Eprint
  {http://arxiv.org/abs/2007.08998}{arXiv:2007.08998 [astro-ph.CO]}\BibitemShut
  {NoStop}%
\bibitem [{\citenamefont {Neveux}\ \emph {et~al.}(2020)\citenamefont {Neveux}
  \emph {et~al.}}]{Neveux:2020voa}%
  \BibitemOpen
  \bibfield  {author} {\bibinfo {author} {\bibfnamefont {R.}~\bibnamefont
  {Neveux}} \emph {et~al.},\ }\bibfield  {title} {\enquote {\bibinfo {title}
  {{The completed SDSS-IV extended Baryon Oscillation Spectroscopic Survey: BAO
  and RSD measurements from the anisotropic power spectrum of the quasar sample
  between redshift 0.8 and 2.2}},}\ }\href {\doibase 10.1093/mnras/staa2780}
  {\bibfield  {journal} {\bibinfo  {journal} {Mon. Not. Roy. Astron. Soc.}\
  }\textbf {\bibinfo {volume} {499}},\ \bibinfo {pages} {210} (\bibinfo {year}
  {2020})},\ \Eprint {http://arxiv.org/abs/2007.08999}{arXiv:2007.08999
  [astro-ph.CO]}\BibitemShut {NoStop}%
\bibitem [{\citenamefont {Tamone}\ \emph {et~al.}(2020)\citenamefont {Tamone}
  \emph {et~al.}}]{Tamone:2020qrl}%
  \BibitemOpen
  \bibfield  {author} {\bibinfo {author} {\bibfnamefont {A.}~\bibnamefont
  {Tamone}} \emph {et~al.},\ }\bibfield  {title} {\enquote {\bibinfo {title}
  {{The Completed SDSS-IV extended Baryon Oscillation Spectroscopic Survey:
  Growth rate of structure measurement from anisotropic clustering analysis in
  configuration space between redshift 0.6 and 1.1 for the Emission Line Galaxy
  sample}},}\ }\href {\doibase 10.1093/mnras/staa3050} {\bibfield  {journal}
  {\bibinfo  {journal} {Mon. Not. Roy. Astron. Soc.}\ }\textbf {\bibinfo
  {volume} {499}},\ \bibinfo {pages} {5527} (\bibinfo {year} {2020})},\ \Eprint
  {http://arxiv.org/abs/2007.09009}{arXiv:2007.09009 [astro-ph.CO]}\BibitemShut
  {NoStop}%
\bibitem [{\citenamefont {de~Mattia}\ \emph {et~al.}(2021)\citenamefont
  {de~Mattia} \emph {et~al.}}]{deMattia:2020fkb}%
  \BibitemOpen
  \bibfield  {author} {\bibinfo {author} {\bibfnamefont {A.}~\bibnamefont
  {de~Mattia}} \emph {et~al.},\ }\bibfield  {title} {\enquote {\bibinfo {title}
  {{The Completed SDSS-IV extended Baryon Oscillation Spectroscopic Survey:
  measurement of the BAO and growth rate of structure of the emission line
  galaxy sample from the anisotropic power spectrum between redshift 0.6 and
  1.1}},}\ }\href {\doibase 10.1093/mnras/staa3891} {\bibfield  {journal}
  {\bibinfo  {journal} {Mon. Not. Roy. Astron. Soc.}\ }\textbf {\bibinfo
  {volume} {501}},\ \bibinfo {pages} {5616} (\bibinfo {year} {2021})},\ \Eprint
  {http://arxiv.org/abs/2007.09008}{arXiv:2007.09008 [astro-ph.CO]}\BibitemShut
  {NoStop}%
\bibitem [{\citenamefont {Nadathur}\ \emph {et~al.}(2020)\citenamefont
  {Nadathur} \emph {et~al.}}]{Nadathur:2020vld}%
  \BibitemOpen
  \bibfield  {author} {\bibinfo {author} {\bibfnamefont {S.}~\bibnamefont
  {Nadathur}} \emph {et~al.},\ }\bibfield  {title} {\enquote {\bibinfo {title}
  {{The completed SDSS-IV extended baryon oscillation spectroscopic survey:
  geometry and growth from the anisotropic void\textendash{}galaxy correlation
  function in the luminous red galaxy sample}},}\ }\href {\doibase
  10.1093/mnras/staa3074} {\bibfield  {journal} {\bibinfo  {journal} {Mon. Not.
  Roy. Astron. Soc.}\ }\textbf {\bibinfo {volume} {499}},\ \bibinfo {pages}
  {4140} (\bibinfo {year} {2020})},\ \bibinfo {note} {[Erratum:
  Mon.Not.Roy.Astron.Soc. 516, 2936--2937 (2022)]},\ \Eprint
  {http://arxiv.org/abs/2008.06060}{arXiv:2008.06060 [astro-ph.CO]}\BibitemShut
  {NoStop}%
\bibitem [{\citenamefont {du~Mas~des Bourboux}\ \emph
  {et~al.}(2020)\citenamefont {du~Mas~des Bourboux} \emph
  {et~al.}}]{duMasdesBourboux:2020pck}%
  \BibitemOpen
  \bibfield  {author} {\bibinfo {author} {\bibfnamefont {H.}~\bibnamefont
  {du~Mas~des Bourboux}} \emph {et~al.},\ }\bibfield  {title} {\enquote
  {\bibinfo {title} {{The Completed SDSS-IV Extended Baryon Oscillation
  Spectroscopic Survey: Baryon Acoustic Oscillations with Ly\ensuremath{\alpha}
  Forests}},}\ }\href {\doibase 10.3847/1538-4357/abb085} {\bibfield  {journal}
  {\bibinfo  {journal} {Astrophys. J.}\ }\textbf {\bibinfo {volume} {901}},\
  \bibinfo {pages} {153} (\bibinfo {year} {2020})},\ \Eprint
  {http://arxiv.org/abs/2007.08995}{arXiv:2007.08995 [astro-ph.CO]}\BibitemShut
  {NoStop}%
\end{thebibliography}%

\end{document}